\documentclass[10pt,onecolumn]{article}
\usepackage[utf8]{inputenc}
\usepackage[english]{babel}
\usepackage{graphicx}
\usepackage{natbib}

\usepackage{authblk}

\title{A Thermal Plume Model for the Martian Convective Boundary Layer}
\author[1]{Arnaud Cola\"itis}
\author[1]{Aymeric Spiga\thanks{Corresponding author: aymeric.spiga@upmc.fr}}
\author[1]{Fr\'ed\'eric Hourdin}
\author[1]{Catherine Rio}
\author[1]{Fran\c cois Forget}
\author[1]{Ehouarn Millour}
\affil[1]{\normalsize Laboratoire de M\'et\'eorologie Dynamique (LMD), Universit\'e Pierre et Marie Curie (UPMC), Institut Pierre Simon Laplace (IPSL), Centre National de la Recherche Scientifique (CNRS), Paris, France}

\date{}

\begin{document}
\maketitle

\emph{ The Martian Planetary Boundary Layer [PBL] is a crucial component of the Martian climate system. Global Climate Models [GCMs] and Mesoscale Models [MMs] lack the resolution to predict PBL mixing which is therefore parameterized. Here we propose to adapt the ``thermal plume" model, recently developed for Earth climate modeling, to Martian GCMs, MMs, and single-column models. The aim of this physically-based parameterization is to represent the effect of organized turbulent structures (updrafts and downdrafts) on the daytime PBL transport, as it is resolved in Large-Eddy Simulations [LESs]. We find that the terrestrial thermal plume model needs to be modified to satisfyingly account for deep turbulent plumes found in the Martian convective PBL. Our Martian thermal plume model qualitatively and quantitatively reproduces the thermal structure of the daytime PBL on Mars: superadiabatic near-surface layer, mixing layer, and overshoot region at PBL top. This model is coupled to surface layer parameterizations taking into account stability and turbulent gustiness to calculate surface-atmosphere fluxes. Those new parameterizations for the surface and mixed layers are validated against near-surface lander measurements. Using a thermal plume model moreover enables a first order estimation of key turbulent quantities (e.g. PBL height, convective plume velocity) in Martian GCMs and MMs without having to run costly LESs.
 }

\section{Introduction}

The exploration of the Martian environment yields many examples of planetary boundary layer [PBL] phenomena commonly encountered on Earth: convective cloud streets \citep{Mali:01}, dust devils \citep[i.e. dusty convective vortices, see][for a review]{Balm:06}, afternoon growth of the mixing layer \citep{Hins:08} associated with turbulent fluctuations of near-surface temperature \citep{Smit:06}, nighttime stable conditions with low-level jets \citep{Savi:93}. The Martian environment can be seen as a large dusty desert in which PBL dynamics is more extreme than on Earth. Owing to the thin~$\mathrm{CO}_2$ atmosphere and low thermal inertia of the surface, the Martian PBL is radiatively controlled and undergoes a strong diurnal cycle with temperature gradients in the surface layer [SL] following superadiabatic regimes in daytime and ultra-stable regimes in nighttime \citep[e.g.][]{Scho:97,Smit:06}. In the afternoon, the mixed layer is sometimes almost as deep as one atmospheric scale height ($\sim$ 10 km) \citep{Hins:08} and, in many cases, deeper than the afternoon PBL in most regions on Earth \citep{Spig:11pss}. The Martian daytime PBL is also characterized by the negligible role played by moist processes and, conversely, the crucial role played by the absorption of infrared emission from the surface by CO$_2$ and, to a lesser extent, dust \citep{Habe:93pbl,Savi:99}.

The existing three-dimensional non-hydrostatic mesoscale models for the Martian atmosphere can be used to assess Martian PBL dynamics in daytime \citep{Toig:03,Mich:04,Rich:07,Tyle:08,Spig:10bl}. In so-called Large-Eddy Simulations [LESs], the grid spacing is lowered to a few tens of meters so as to resolve the largest turbulent eddies, responsible for most of the energy transport within the PBL \citep{Lill:62}. LESs have demonstrated that, from late morning to sunset, PBL dynamics associated with superadiabatic near-surface temperature gradients comprise powerful narrow updrafts with vertical velocities of~$10-20$~m~s$^{-1}$ and broad downdrafts with vertical velocities of~$5-10$~m~s$^{-1}$, organized in a polygonal cellular structure \citep{Mich:04,Spig:09}. Recent LESs reproduce the regional variability of PBL depth revealed through observations and dominated on Mars by radiative forcing inside the boundary layer \citep{Spig:10bl}.

\begin{figure*}
\begin{center}
\noindent\includegraphics[width=\columnwidth]{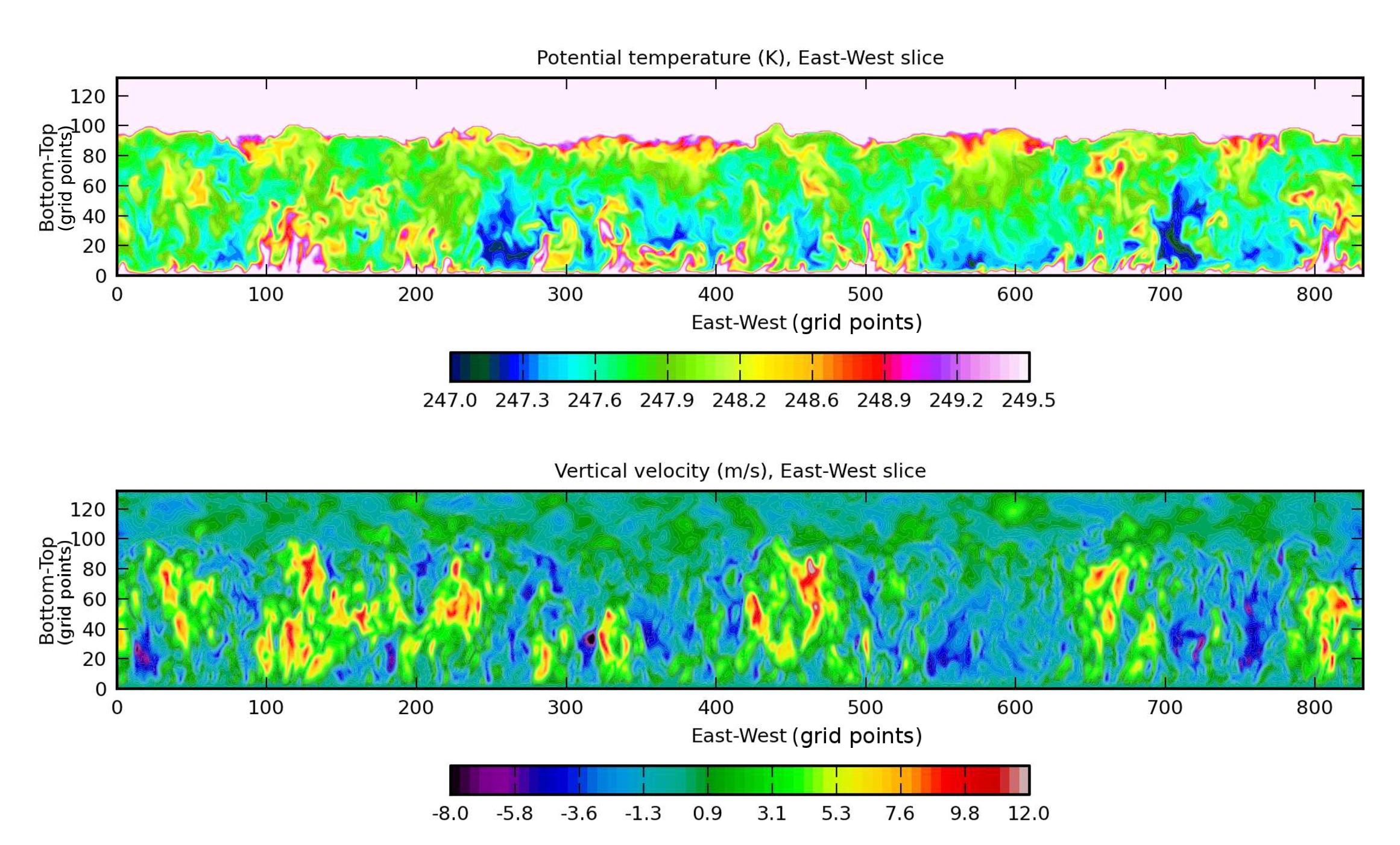}
\caption{LES results for the Martian convective boundary layer. Slices of potential temperature (K) [top] and vertical velocity ($\mathrm{m.s}^{-1}$) [bottom]. Results are from simulation C.large, taken at local time 15:00 along the West-East direction (aligned with the background wind). Horizontal resolution for this simulation is 100m between grid points so that total domain extent is 83km. Vertical resolution is about 75m (except in the first few layers where the mesh is refined for the surface layer).}
\label{fig:ncviewdowndrafts1}
\end{center}
\end{figure*}

Convective plumes, i.e. the largest eddies resolved in LESs, are named non-local turbulence, or organized turbulence: e.g. updrafts entrain air from the surface layer and detrain it at several kilometers above the ground in daytime (cf. Figure~\ref{fig:ncviewdowndrafts1}). Conversely, local turbulence refers to turbulent motions which do not induce vertical transport over a significant fraction of the PBL depth. Both large eddies, resolved by LESs, and small eddies, unresolved by LESs, contribute to local turbulent mixing. In global climate models [GCMs], mesoscale models [MMs], and single-column models [SCMs], all three kinds of turbulent structures in the PBL (non-local large eddies, local large eddies, small-scale eddies) are left unresolved, and hence must be parameterized.

Parameterizing PBL vertical transport in GCMs and MMs is a key element to accurately predict the large-scale and regional variability of winds and temperature, volatile mixing and surface-atmosphere interactions [e.g. dust lifting]. Local turbulence is usually parameterized by turbulent closure schemes, where mixing is obtained through local diffusion processes \citep[][for a review]{Mell:82}. A remaining crucial issue is how to model non-local turbulence, which participates significantly to the total turbulent mixing budget of the daytime convective PBL. 

The crudest parameterization for non-local turbulence consists in replacing an unstable PBL profile by its neutral equivalent while conserving mass and energy (convective adjustment). This approach strongly underestimates daytime near-surface temperatures in Martian GCMs and MMs \citep{Rafk:03adj,Spig:09} because the radiatively-controlled superadiabatic layers in the first hundreds of meters above the Martian surface cannot be reproduced. More sophisticated techniques, inherited from terrestrial modeling, were adopted instead \citep{Tyle:02,Rich:07}. A widely used method in Martian models is to parameterize non-local turbulence by adding a countergradient term in local turbulent diffusion schemes \citep{Troe:86}. 

``Thermal plume models" have been recently developed in terrestrial GCMs and MMs to parameterize non-local large eddies and the resulting transport in the PBL \citep{Hour:02,Soar:04,Rio:08}. Their name reflects their aim to model non-local transport by describing convective plumes in the daytime PBL more explicitly than counter-gradient schemes. In other words, thermal plume models attempt not only to parameterize the mixed layer within the PBL, but also the transport processes responsible for it. These schemes, also named mass flux PBL schemes, are now routinely used in several terrestrial GCMs \citep{Teix:11,Hour:12}, and were shown to yield a better representation of water vapor transport from the surface to the free atmosphere.

Using a thermal plume model in Martian GCMs and MMs has a great potential for improving the representation of PBL processes in these models:
\begin{enumerate}
\item The vigorous daytime PBL mixing on Mars by strong updrafts and downdrafts makes the thermal plume model especially relevant to the Martian environment. Adapting this model to Mars in turn offers the possibility to test it, and possibly improve it, by using it in an extra-terrestrial desert, devoid of any significant moist processes, and prone to original and extreme PBL dynamics.
\item \cite{Tyle:08} noticed the daytime PBL depth parameterized in Martian MMs is significantly underestimated compared to LESs. A possible explanation might be that countergradient schemes are not accurate enough to represent non-local turbulence in the strongly convective Martian PBL.
\item Thermal plume models allow for key PBL variables (e.g. updraft/downdraft speeds) to be estimated to first order in GCMs and MMs. Furthermore, a proxy for near-surface gustiness can be obtained from thermal plume models, which permits a more realistic representation of surface-atmosphere fluxes.
\end{enumerate}

Here we report the first adaptation of a thermal plume model to simulate PBL dynamics and mixing in Martian GCMs, MMs, and SCMs. We describe in section \ref{sec:sampling} the methods used to analyze and sample the convective structures in LES of the Martian daytime PBL. The thermal plume model for Mars is formulated in sections \ref{sec:Martianmodel} and \ref{sec:updraftanddowndraft}. An improved surface layer model for Mars, coupled to parameterized turbulent gustiness in the thermal plume model, is proposed in section \ref{sec:sl}. In section~\ref{sec:resultsthermals}, we test our new PBL parameterizations, discuss their performances compared to LES results, and validate their predictions with in-situ data on Mars. In all sections, symbols are defined at their first appearance in the paper and listed within the supplementary material.

\section{Large-Eddy Simulations of Martian daytime PBL convection}
\label{sec:sampling}

A prerequisite for the adaptation of the thermal plume model to Mars is a characterization of the non-local convective plumes in the Martian daytime PBL. This is done through LES and sampling techniques adapted for Mars.

\subsection{Methodology and results}

Martian LES are carried out with the Laboratoire de M\'et\'eorologie Dynamique [LMD] Martian Mesoscale Model by \cite{Spig:09}, based on the Weather Research Forecast dynamical core \citep{Skam:08} and its adaptations for LES \citep{Moen:07}. We adopt similar settings and physical parameterizations as in \cite{Spig:10bl} (see also \cite{Forg:99} and \cite{Made:11} for parameterizations used in LMD models, namely radiative transfer). We use a~$101 \times 101 \times 201$ grid with a horizontal resolution of~$100$~m and a vertical resolution of~$75$~m. \cite{Spig:10bl} showed that these LES correctly represent the dynamics of the daytime convective PBL. Finer resolution LESs are performed in this paper for verification purposes and a better sampling of convective structures. 

LESs are performed over a large range of environmental conditions relevant to Mars in order to assess the relevance of the thermal plume approach. Those cases, summarized in tables~\ref{tab:1} and~\ref{tab:2}, are inspired both by reference cases for which observations are available (radio-occultations by \cite{Hins:08} compared to LES by \cite{Spig:10bl}) and extreme cases for PBL convection obtained e.g. by setting low albedo, thermal inertia, or surface pressure. All simulations are initialized at local time~06:00 (before convection becomes active) using the climatologies of the Martian LMD GCM \citep{Forg:99} available in the ``Mars Climate Database" \citep[MCD,][]{Mill:08mcd}. A background wind of~$10$~m~s$^{-1}$ is prescribed to represent typical synoptic / regional circulations on Mars (the indicated value is valid within the mixed layer; in the surface layer, momentum mixing and near-surface friction produce background wind speed lower than prescribed). To improve the characterization of PBL non-local turbulence for the thermal plume model, namely updraft velocity, entrainment and detrainment rates, we also ran a more computationally challenging LES over a~80 km~$\times$~80~km domain for case~C. About 30 updrafts are featured in this LES, compared to 1-2 with reference settings.

%%%%%%%%%%%%%%%%%
\begin{table*}
%\begin{tabular*}{1.0\mathrmwidth}{@{\extracolsep{\fill}} l c c c c c c c c r}
\begin{tabular}{ l c c c c c c c c r }
\hline
    &~$L_s$ &~$\varphi$ &~$\lambda$ &~$\mathcal{T}$ &~$\mathcal{A}$ &~$T_s$ &~$h$ &~$p_s$\\
      & ($^{\circ}$) & ($^{\circ}$N) & ($^{\circ}$E) & (tiu) &   & (K, at 6:00) & (km) & ($P_a$, at 6:00) &~$\tau$\\
   \hline
   A & 47.1 & 21.8 & 205.0 & 55 & 0.27 & 167.32 & -3.9 & 857.0 & 0.5\\
   C & 52.1 & 12.3 & 237.2 & 60 & 0.30 & 161.15 & +2.5 & 480.7 & 0.5\\
   I  & 47.8 & 20.6 & 74.0 & 300 & 0.13 & 192.56 & -0.5 & 629.0 & 0.5\\
   Z & 67.0 & -10.2 & 236.6 & 42 & 0.28 & 153.04 & + 8.4 & 265.6 & 0.5\\
   Exo & 244 & -1.82 & -6.15 & 238 & 0.205 & 197.3 & -1.4 & 717.9 & 0.5\\
   E & 0. & 0. & 0. & 50 & 0.10 & 186.26 & -1.5 & 670.0 & 0.05\\
 \hline
 \end{tabular}
   \caption{ Settings of reference LES for the design of the Martian thermal plume model. $L_s$ is the solar longitude,~$\varphi$ is the north latitude,~$\lambda$ is the east longitude,~$\mathcal{T}$ is the surface thermal inertia in J~m$^{-2}$~K$^{-1}$~s$^{-1/2}$ [tiu],~$\mathcal{A}$ is the surface albedo,~$T_s$ is the surface temperature at local time 6:00 (corresponding to starting time for simulations),~$h$ is the altitude with respect to the MOLA reference, ~$p_s$ is the surface pressure at local time 6:00 and~$\tau$ is the dust opacity ($\tau = 0.5$ corresponds to moderately dusty conditions on Mars). Cases A, C, I, and Z are inspired by \cite{Hins:08} observations and follow their naming convention (see also \cite{Spig:10bl}). Case Exo [Exomars] is a simulation at Meridiani Planum (reference site for Exomars preparatory studies). Case E is an ``extreme" convection case obtained by assuming a clear atmosphere and low thermal inertia and albedo.}
\label{tab:1}
\end{table*}

\begin{table*}
%\begin{tabular*}{1.0\mathrmwidth}{@{\extracolsep{\fill}} l c c c c c c r}
\begin{tabular}{ l c c c c c c r }
\hline
    & Grid &~$\mathrm{dx}=\mathrm{dy}$ &~$\Delta x = \Delta y$ &~$\mathrm{dz}$ &~$\Delta z$ &~$\tau$ &~$u$\\
    &~$N_x$ x~$N_y$ x~$N_z$ & (m) & (km) & (m) &  (km) &  & (m/s)\\
   \hline
   default & $101 \times 101 \times 201$ & 100 & 10.0 & 75 & 15.0 & 0.5 & 10.0\\
   .hr & $257 \times 257 \times 301$ & 50 & 12.8 & 50 & 15.0 & - & -\\
   .large & $833 \times 833 \times 133$ & 100 & 83.2 & 75 & 10 & - & -\\
   .t2 & - & - & - & - & - & 2. & -\\
   .w30 & - & - & - & - & - & - & 30.0\\
 \hline
 \end{tabular}
  \caption{ Additional cases associated with reference cases described in Table \ref{tab:1}.~dx and~dy (resp.~$\Delta$x and~$\Delta$y) are horizontal resolution (resp. domain size) in the West-East and South-North directions,~$u$ the background wind in the West-East direction and dz is the vertical resolution above the surface layer. '-' indicates default value. Changes in settings are mostly applied to cases A and C. }
\label{tab:2}
\end{table*}

%%%%%%%%%%%%%%%%%

Figure~\ref{fig:ncviewdowndrafts1} illustrates the typical non-local large eddies which develop in the Martian convective PBL in daytime in this reference simulation. The observed ``flame-like" structures are associated with strong vertical velocities and positive potential temperature anomalies compared to slab averages (by slab we mean a domain-wide square at a given altitude or model level). These correspond to updrafts and exhibit large buoyancies. Cold areas can also be identified, mostly corresponding to areas of negative vertical velocities. Plumes detrain mainly at the top of the boundary layer, depositing the remainder of the advected heat there.

\subsection{Sampling non-local structures}
\label{subsec:updraftselection}

A first step towards parameterizing the non-local turbulence resolved by LESs is to sample organized structures, namely updrafts and downdrafts. Figure~\ref{fig:ncviewdowndrafts1} emphasizes the difficulty in choosing between what can be considered as part of an updraft and what can be considered as the environment. The distinction between the two must be formulated adequately, as the variations of the mass flux of the plume, hence vertical mixing, are directly connected to entrainment and detrainment rates. 

\begin{figure*}
\begin{center}
		\noindent\includegraphics[width=\columnwidth]{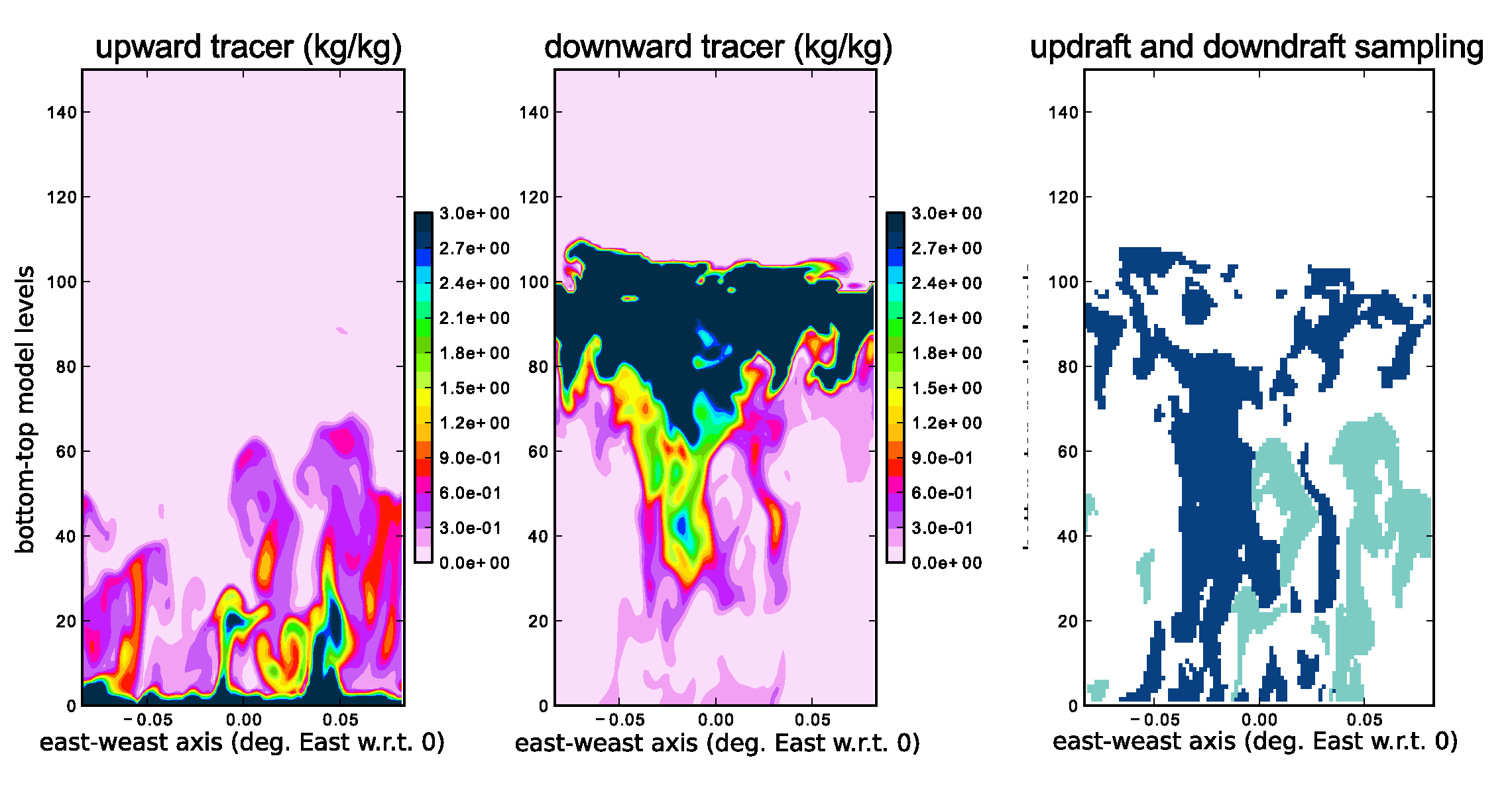}
		\caption{ Using LES to sample organized convective structures (case I at local time 15:20). Slices of upward [left] and downward [center] decaying tracer concentrations ($\mathrm{kg.kg}^{-1}$) respectively emitted at the surface and in the PBL top. Results of the conditional sampling selection are shown in the right panel where dark blue is downdraft sampling and light green is updraft sampling.}
		\label{fig:sampling}
\end{center}
\end{figure*}

Several criteria have been proposed to sample organized structures in terrestrial LESs, mostly based on quantities related to water. For Mars, we use the tracer-based conditional sampling formulation proposed by \citet{Couv:10}, which is not water-based and shows satisfactory results with respect to other methods in Earth LESs. A tracer is emitted in the first layer of the LES model, with a decaying concentration whose half-life is determined by the time it would take a particle to reach the top of the PBL. Typical timescales for this migration on Mars are between 5 and 10 minutes, which is much less than on Earth (typically 20 to 60 minutes). Figure~\ref{fig:sampling} shows the concentration of such a decaying tracer emitted in the surface layer, with a 600 seconds half-life.

A point~$M(x,y,z,t)$ in the LES grid is assumed to belong to an updraft if it satisfies:

\begin{equation}
q'(M) > \gamma \mathrm{\ max}(\sigma_{q},\sigma_{\mathrm{min}}) \quad \mathrm{and} \quad w(M) > 0
\label{eq:CSU}
\end{equation}

where~$q'$ is the tracer anomaly with respect to the slab average,~$w$ is the vertical velocity at the chosen point,~$\gamma$ is a scale factor (chosen to be one),~$\sigma_q$ is the standard deviation of tracer concentration at the corresponding level and~$\sigma_{\mathrm{min}}$ is a minimum standard deviation designed to avoid selecting too many points in well-mixed layers. The value of~$\sigma_{\mathrm{min}}$ proves to be especially important close to the PBL top, where turbulence is strong and updrafts detrain. The minimum standard deviation proposed by \citet{Couv:10} is:

\begin{equation}
\sigma_{\mathrm{min}} = \frac{\sigma_0}{z}\int_0^z{\sigma_q(k)\,\partial k}
\label{eq:couvr}
\end{equation}

where the scaling factor~$\sigma_0$ is 0.05. Tests and comparisons to LESs suggest that a value of~$\sigma_0=0.2$ is more suited to Martian convection, as lower values yield a too large fractional coverage in the detraining zone at the PBL top. The updraft selection is considered to be satisfactory if it maximizes the heat and mass flux of the updraft for a minimum fraction coverage. 

We choose to apply the same kind of sampling for downdrafts, except our decaying tracer is not emitted at the surface but in the putative layer from which downdrafts originate. This source for downdrafts is identified in LESs as being near the PBL top (Figure~\ref{fig:sampling}). We compute for each timestep from a first LES run the PBL height~$z_i$, defined as the altitude at which the mean vertical velocity in the plumes reaches zero. The downdraft tracer is emitted in a second LES run at the predetermined values of~$z_i$. To avoid cumbersome tracer repartitions in the PBL, tracer emission is only activated when the PBL height is almost stationary, between 12:00 and 17:00. 

A point~$M(x,y,z,t)$ in the LES grid is assumed to belong to a downdraft if it satisfies:

\begin{equation}
q'(M) > \gamma \mathrm{\ max}(\sigma_{q},\sigma_{\mathrm{min}}') \quad \mathrm{and} \quad w(M) < 0 \nonumber
\label{eq:CSD}
\end{equation}

where~$\sigma_{\mathrm{min}}'$ is defined as in equation~\ref{eq:couvr} except for integration boundaries~$[z_i,z]$. To ensure that only the most prominent downward structures are sampled, the scale factor~$\gamma$ for tracer anomaly in equation~(\ref{eq:CSU}) is increased from 1 to 1.5. 

\begin{figure}
\begin{center}
\noindent\includegraphics[width=0.5\columnwidth]{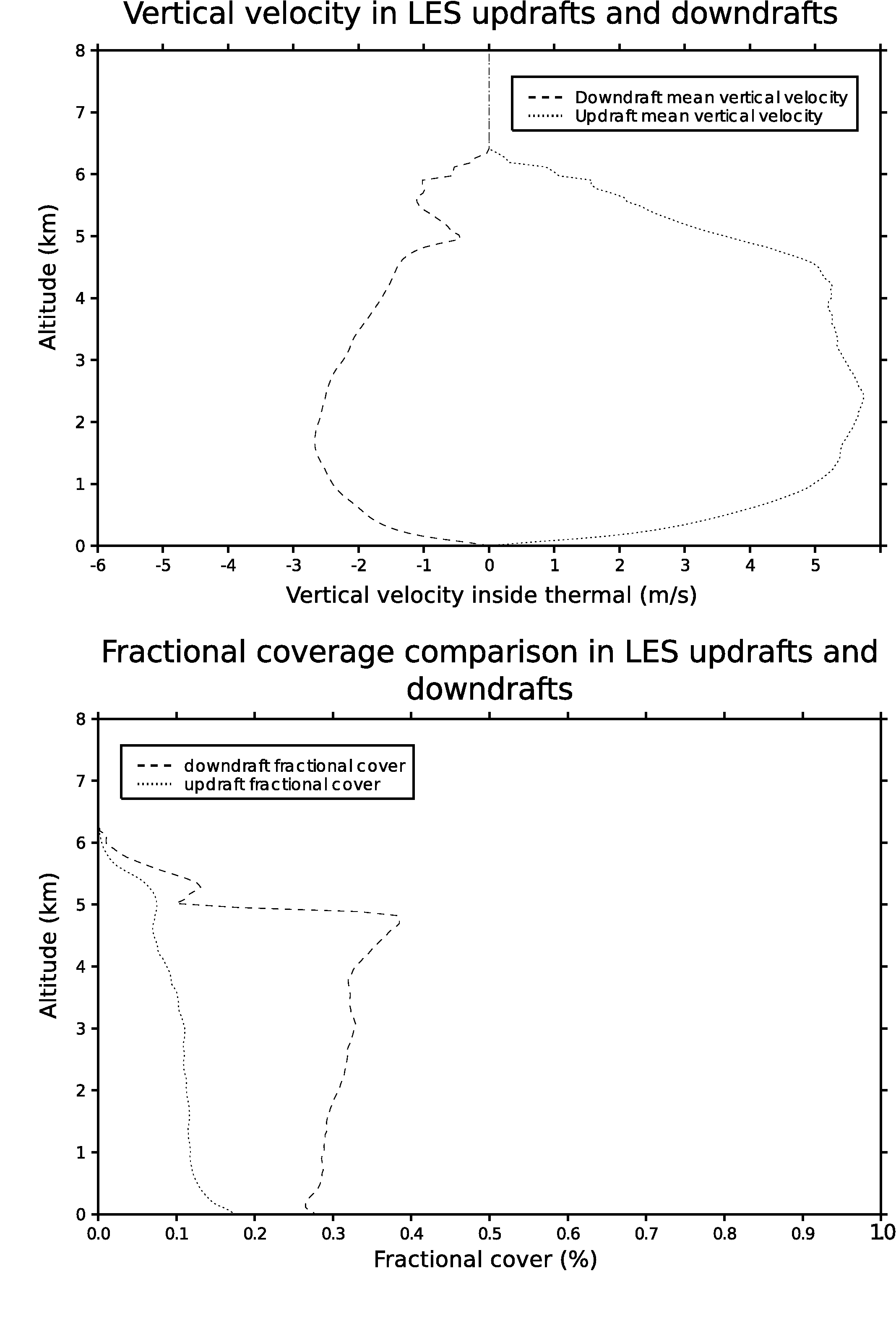}
\caption{ Structure-averaged vertical velocities [top] and fractional coverages [bottom] for updrafts (dotted lines) and downdrafts (dashed lines) using the conditional sampling technique on case I, at local time 15:00, averaged over a 925 s window.}
\label{fig:alphaVCSD}
\end{center}
\end{figure}

The result of the conditional sampling criterion on both updrafts and downdrafts is illustrated in Figure~\ref{fig:sampling}. Structure-averaged velocities for updrafts and downdrafts are shown for case I in Figure~\ref{fig:alphaVCSD}, alongside fractional coverages. Downdraft velocities reach about half of the maximum updraft speed and cover a factor of two to three larger area than updrafts. In the particular example of Figure~\ref{fig:alphaVCSD}, downdrafts can be considered to originate between 4 and 5 km. As downward air accelerates, the plume becomes thinner. These profiles do appear like inverted updrafts starting in the inversion layer. Analyzing the downdrafts predicted in case C.large yields slightly different results than with smaller domains: fractional coverages of about~$20\%$ and vertical velocities about half of the updraft velocities.

\section{A thermal plume model for Mars}
\label{sec:Martianmodel}

\subsection{Two-column formulation}
\label{subsec:earth}

Our thermal plume model for Mars is based on a modified version of the terrestrial model by \citet{Hour:02} and \citet{Rio:08}. A thermal plume model describes a plume of air rising in the PBL through the effect of its buoyancy, fed by horizontal winds in the turbulent surface layer \citep[Figure 1 in][]{Rio:08}. A model column, corresponding to a single column in GCMs and MMs' grids, is separated into two fictive sub-columns: updraft and environment. This decomposition enables the expression of a conserved variable~$\phi$ (potential temperature, momentum, tracer concentration, \ldots) separately for these two sub-columns:

\begin{equation}
\overline{\phi} = \alpha \, \phi_u + (1-\alpha) \, \phi_e
\label{eq:2phi}
\end{equation}

where~$\alpha$ is the updraft fractional coverage, and subscripts~$u$ and~$e$ respectively stand for updraft-averaged and environment-averaged values. Overbar quantities denote slab-averaged quantities for one model column, equivalent to slab-averaged quantities over a LES domain (the extent of which is, at best, about the grid spacing in GCMs or MMs). 

At each level, the thermal plume exchanges air with the environment through entrainment and detrainment. It rises while its buoyancy is positive, and overshoots when its buoyancy is negative. In this process, air from lower levels rises, and detrains in the environment at higher levels. This upward transport by buoyant plumes is coupled through PBL convective cells with downward compensating motions. On Earth, updrafts usually cover a small fraction of the area of a GCM or MM grid mesh: compensatory subsidences are slower and have larger fractional coverages than updrafts. 

Following e.g. \cite{Sieb:95}, the vertical turbulent transport~$\overline{w'\phi'}$ of a scalar~$\phi$ can be formulated as: 

\begin{eqnarray}
 \overline{w'\phi'}  & = & \alpha \, \overline{w'\phi'}^u + (1-\alpha) \, \overline{w'\phi'}^e \\
                     &   & + \alpha \, (1-\alpha) \, (w_u - w_e) \, (\phi_u - \phi_e)
\label{eq:2full}
\end{eqnarray}

where the prime symbol is the deviation from slab-average, $\overline{w'\phi'}^u$ is the updraft-averaged product of deviations from the updraft average and~$\overline{w'\phi'}^e$ the environment-averaged product of deviations from the environment average. The first two terms in equation~\ref{eq:2full} represent the contribution of local large eddies inside the thermal plume and the environment respectively. The third term (named organized turbulence) accounts for the transport by non-local large eddies, i.e. upward plumes and compensating subsidence. 

Terrestrial LESs have shown that the contribution to PBL transport of local eddies within thermal plumes is not significant. Hence, in Earth thermal plume models, the first term in equation~\ref{eq:2full} is neglected. In most cases, the second term is also found to be negligible and the vertical turbulent flux of~$\phi$ simplifies to the third term. Assuming that the updraft fractional coverage~$\alpha$ is small, ~$\alpha^2$ terms are negligible and equation~\ref{eq:2phi} yields~$w_e \simeq \overline{w}$. Finally, equation~\ref{eq:2full} simplifies to:

\begin{equation}
\overline{w'\phi'} = \frac{F_u}{\rho} (\phi_u - \phi_e)
\label{eq:turbulenceE}
\end{equation}

where~$F_u = \alpha \, \rho \, w_u$ is the updraft mass flux.

\subsection{Three-column formulation with downdrafts}
\label{subsec:reassessment}

\cite{Couv:07} showed that the turbulent transport by downdrafts can sometimes be significant, when ``dry tongues'' form in the convective PBL on Earth. This questions the validity of the simplified equation~\ref{eq:turbulenceE} used in thermal plume models. This limitation appears all the more critical in the Martian convective PBL, given the intensity of downdrafts observed in LESs (Figure~\ref{fig:ncviewdowndrafts1}). The contribution of these downdrafts to non-local mixing in the daytime PBL must be taken into account. 

In our Martian thermal plume model, we add a downdraft sub-column to the updraft and environment sub-columns commonly considered in terrestrial thermal plume models \citep[see discussions in section 4 of][]{Sieb:95}. This decomposition along 3 sub-columns is inspired by the \citet{Tied:89} deep convection mass flux scheme. This is in line with the fact that PBL convection, also named ``shallow" convection on Earth, is actually not so shallow on Mars where non-local PBL transport can reach about one atmospheric scale height.

The decomposition of the flux of a scalar~$\phi$ in equation~\ref{eq:2full} now reads:

\begin{eqnarray}
\overline{w'\phi'} = &   \alpha_u \overline{w'\phi'}^u + \alpha_d \overline{w'\phi'}^d + (1 - \alpha_u - \alpha_d) \overline{w'\phi'}^e \\
                     & + \alpha_u(w_u - \overline{w})(\phi_u - \overline{\phi}) + \alpha_d(w_d - \overline{w})(\phi_d - \overline{\phi}) \\
                     & + (1 - \alpha_u - \alpha_d)(w_e - \overline{w})(\phi_e - \overline{\phi})
\end{eqnarray}

where subscript~$d$ denotes downdraft values. The first three terms represent local turbulence within the updraft, downdraft and environment. The last three terms are the turbulence arising from non-local (or organized) structures: the first one is associated with updrafts, the second one with downdrafts and the last one with the environment. As in section~\ref{subsec:earth}, we assume local turbulence (the first two terms) can be neglected, and turbulence in the environment will be parameterized by diffusion schemes. This leads to a simplified expression for the turbulent flux of~$\phi$:

\begin{eqnarray}
\overline{w'\phi'} = & \alpha_u(w_u - \overline{w})(\phi_u - \overline{\phi})+\alpha_d(w_d - \overline{w})(\phi_d - \overline{\phi}) \\
                     & + (1 - \alpha_u - \alpha_d)(w_e - \overline{w})(\phi_e - \overline{\phi})
\label{eq:3terms}
\end{eqnarray}

The turbulent kinetic energy [TKE] equation,  where~$\mathrm{TKE} =  0.5 \, (\overline{u'^2}+\overline{v'^2}+\overline{w'^2})$ with horizontal velocities~$u$ and~$v$, can be defined similarly (see supplementary material section 2). 

Equation~\ref{eq:3terms} can be further simplified by comparing environmental values of vertical velocity~$w$ and potential temperature~$\theta$ to LESs slab-averaged values.~$\theta_e = \overline{\theta}$ is a good approximation which actually holds for any conserved variable~$\phi$, since mixing occurs in the same way for all these variables. This yields:

\begin{equation}
\overline{\phi} = \alpha_u\phi_u + \alpha_d\phi_d + (1-\alpha_u-\alpha_d)\phi_e \simeq  \phi_e .
\end{equation}

\noindent $\overline{w}$ corresponds to large-scale and regional vertical motions, which are both slower and less intense than PBL convective motions. To first order, we thus have in equation~\ref{eq:3terms}:~$w_u - \overline{w} \simeq w_u$,~$w_d - \overline{w} \simeq w_d$, and~$w_e - \overline{w} \simeq 0$.

Finally, by introducing a downward mass flux~$F_d = \alpha_d\,\rho\,w_d$, the vertical turbulent transport of~$\phi$ takes the following simple form:

\begin{equation}
\overline{w'\phi'} \simeq \frac{F_u}{\rho}(\phi_u - \overline{\phi}) + \frac{F_d}{\rho}(\phi_d - \overline{\phi})
\label{eq:turbulenceEq}
\end{equation}

This formulation completes equation~\ref{eq:turbulenceE}, and distinguishes updrafts from downdrafts in a convenient way, which allows for using two-column thermal plume models developed for Earth with only minor adaptations. 

\subsection{Comparison of the two formulations}
\label{subsec:Validity}

To compare the two-column and three-column decompositions, we assess in LESs the partitioning of turbulent heat flux and TKE between local and non-local (organized) turbulent structures.

\begin{figure*}
\begin{center}
\noindent\includegraphics[width=\columnwidth]{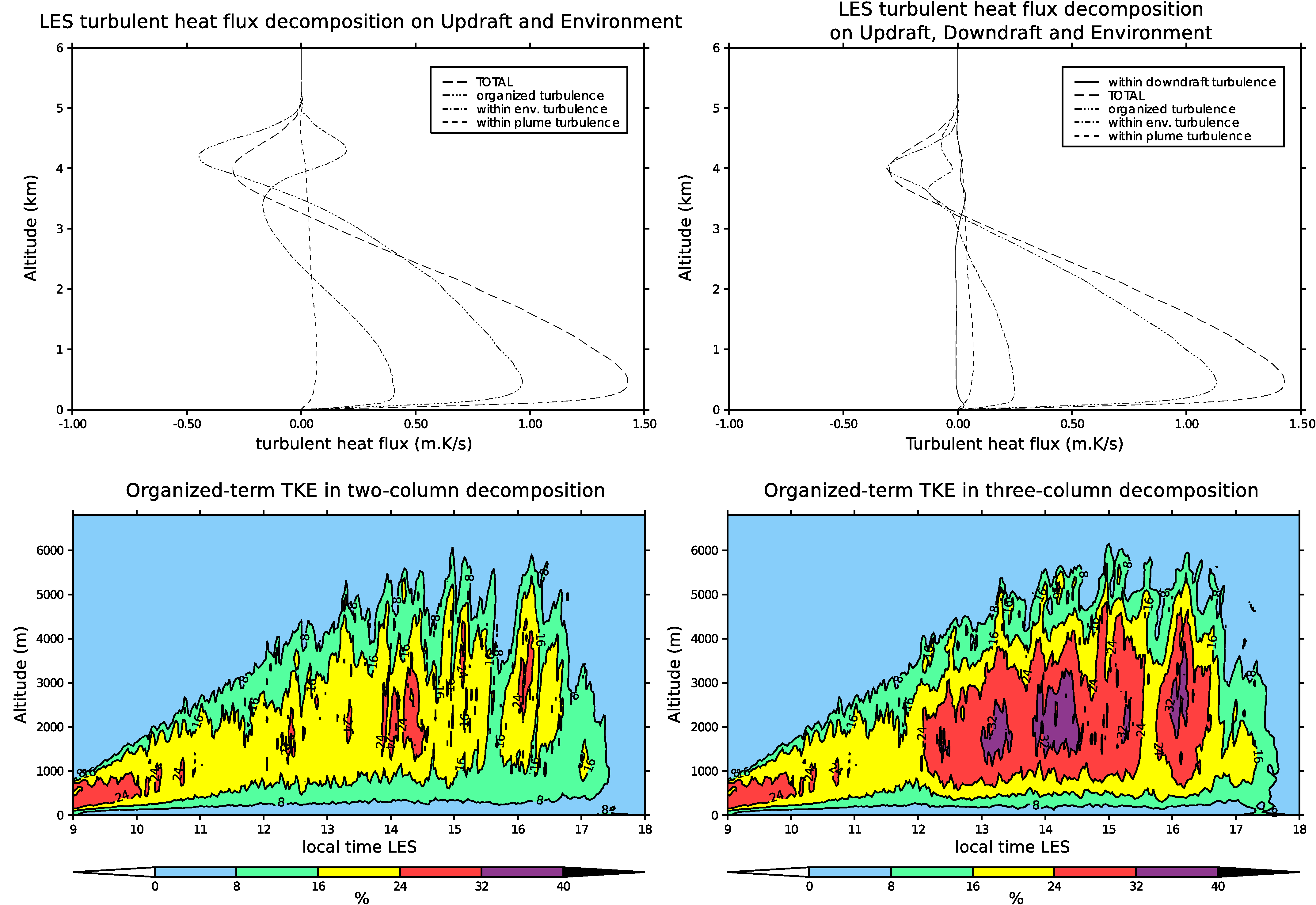}
\caption{ The role of PBL downdrafts on Mars. [Top] Decomposition of the different terms in equation (\ref{eq:2full}) [top-left] and in equation (\ref{eq:3terms}) [top-right]. LES results are from case I, averaged over a 1850-second window centered on local time 13:00 and averaged over the whole domain (slab-average). Long dashed lines represent the total slab-averaged value of~$w'\theta'$. [Bottom] Slab-averaged percentage of resolved TKE associated with organized structures in the terrestrial decomposition [bottom-left] and in the 3 sub-columns decomposition [bottom-right] (results from case I). Computations before 12:00 for the latter figure are conducted assuming that downdrafts have zero fractional coverage, as the downdraft-sampling tracer is only emitted from 12:00 to 17:00. Subgrid-scale TKE from the LES diffusion scheme represents between 10 and 20\% of total TKE in these cases and was added to the environmental part (because these are small eddies) for the computation of TKE percentage associated with organized structures.}
\label{fig:turbuAndTkeUDandUDE}
\end{center}
\end{figure*}

For the two-column decomposition described in section~\ref{subsec:earth}, each term of the turbulent heat flux (i.e. for~$\phi = \theta$ in equation~\ref{eq:2full}) is displayed in Figure~\ref{fig:turbuAndTkeUDandUDE} [top-left] for one typical LES (results are similar for other cases). As is the case on Earth, local fluctuations within Martian updrafts do not contribute significantly to transport. Conversely, turbulence inside the environment contributes to almost a third of the total heat flux. This is due to both strong negative vertical velocities in downdrafts (which can reach up to two thirds of mean updraft velocities on Mars) and resolved local turbulence. Results for turbulent kinetic energy are shown in Figure~\ref{fig:turbuAndTkeUDandUDE} [bottom-left]. According to equation~\ref{eq:2full}, the contribution from non-local large eddies (organized turbulence) is found to represent generally between~$15-25\%$ (peaking locally at~$35\%$) of the TKE resolved through LESs in the mixing layer. This would leave about 80\% of TKE non-parameterized by the thermal plume model.

If this remaining TKE is to be parameterized by a diffusion scheme of the \citet{Mell:74} type, this turbulence should, ideally, only be local. This does not appear realistic on Mars. Figure~\ref{fig:ncviewdowndrafts1} depicts non-local downward structures in the Martian PBL, correlated with broad areas of negative vertical velocities and cold temperatures. Other LES results in the literature have also shown broad downdrafts with large vertical velocities despite temperatures close to the environment along most of the vertical extension of the mixing layer \citep{Mich:04}. In the case of Mars, contrary to the Earth, these structures cannot be part of the~$\overline{w'\phi'}^e$ term and considered as local turbulence. In other words, everything outside the updraft cannot be considered simply as ``environmental" slowly moving air. The same remark could stand for the ``dry tongues" evidenced in the terrestrial PBL by \cite{Couv:07}.

What are the improvements obtained with the three-column decomposition in section~\ref{subsec:reassessment}? The three organized terms for heat flux ($\phi = \theta$ in equation~\ref{eq:3terms}) are grouped into one and plotted in Figure~\ref{fig:turbuAndTkeUDandUDE} [top-right], alongside the three local turbulence terms. This confirms that local turbulence is negligible, and that environmental turbulence can be handled by a local diffusion scheme. The organized turbulence term is found to represent 80\% of the total heat flux. In our three sub-column decomposition, the organized TKE term is generally between~$25-40\%$ of the total resolved TKE in LESs (Figure~\ref{fig:turbuAndTkeUDandUDE} bottom-right; an analysis with case C.large yields similar conclusions). This significant increase in organized TKE, obtained by taking into account non-local downdrafts, confirms the conclusions in section \ref{subsec:reassessment}. In the Martian atmosphere, a part of the environmental turbulence defined through terrestrial decomposition can be seen neither as a slowly compensating subsidence, nor as small eddies. 

\section{Parameterizing updrafts and downdrafts}
\label{sec:updraftanddowndraft}

The next step in developing the thermal plume model is the computation of the values of mass flux~$F$ and variables~$\phi$ for updraft and downdraft sub-columns in equation \ref{eq:turbulenceEq}. 

\subsection{Entrainment and detrainment}

How entrainment and detrainment are parameterized is key to the thermal plume model. The vertical variation of updraft mass-flux~$F_u$ (defined in section~\ref{subsec:earth}) is indeed the difference between entrainment rate~$\epsilon$ and detrainment rate~$\delta$ \citep{Rio:08}. Hence the steady-state conservation equation for a variable~$\phi$ at a given level in a simple updraft / environment decomposition is:

\begin{equation}
\frac{1}{F_u} \, \frac{\partial F_u \, \phi_u}{\partial z} = \epsilon \, \phi - \delta \, \phi_u
\label{eq:conservationphi}
\end{equation}

Formulations used to extract entrainment and detrainment rates~$\epsilon$ and~$\delta$ from LES results are detailed in supplementary material (section 3). 

In equation~\ref{eq:conservationphi}, inherited from terrestrial parameterizations, thermals are assumed to be, at a given time, stationary (i.e. temporal derivative terms in equations detailed in supplementary material section 3 are neglected). Despite the rapid and intense growth of the Martian boundary layer during the day, profiles of~$\epsilon$ and~$\delta$ on Mars indicate that this simplification is still valid on Mars. A steep increase of the contribution of these terms is observed in the surface layer and corresponds to smaller-scale turbulence, represented by the source layer [a prescribed entrainment rate in the surface layer that initiates the thermal plume, the source layer is discussed in the supplementary material section 4]. Above the inversion layer, these terms also become large, consistent with the strong plume detrainment and increase in PBL height.  

\subsection{Vertical velocity equation}

Many distinct formulations for~$\epsilon$ and~$\delta$ are described in Earth literature. In preliminary models, entrainment and detrainment rates were prescribed as being constant with height. More sophisticated parameterizations use schemes where entrainment and detrainment rates also depend on vertical velocity and buoyancy in the plume \citep{Rio:10,Rooy:10}. 

To obtain an equation for the vertical velocity, equation~\ref{eq:conservationphi} is applied to vertical velocity~$w_u$, with an additional term~$\alpha \, \rho \, \Gamma$ that accounts for the lifting and drag forces applied to air parcels within the plume, and the continuity equation in the quasi-Boussinesq approximation is used (see supplementary material section 3). With~$\Omega = w_e / w_u$, this yields:

\begin{equation}
\frac{1}{2}\frac{\partial w_u^2}{\partial z} = - \epsilon \, w_u^2 \, (1-\Omega)  + \Gamma
\label{eq:verticalw}
\end{equation}

where the term~$\rho \, \partial \alpha_u / \partial t$ is neglected. In terrestrial models,~$\Omega$ is assumed to be 0. Nevertheless, because of the strategy we adopt to parameterize downdrafts, we set a small negative value for~$\Omega$, which slightly increases drag induced by entrainment (see section \ref{subsec:downdraft}).

Although realistic expressions exist for~$\Gamma$ \citep{Greg:01}, this term essentially results in a positive contribution to lifting from buoyancy and a negative contribution from drag forces. Following observations on Earth, several authors proposed to reduce the drag term to a term proportional to the square of the vertical velocity \citep{Simp:69,Bret:04}. In our Martian thermal plume model, we parametrize updrafts with the approach adopted by \citet{Rio:10}:

\begin{equation}
\Gamma = a \, B - b \, w_u^2
\label{eq:GammaRio}
\end{equation}

where~$B$ is the buoyancy, $a$ and~$b$ are free parameters. We find that the standard setup~$a=1$ is compliant with Martian LESs which show that~$\Gamma$ is close to the buoyancy profile of the plume (see Figure 1 in supplementary material, we use case C.large to maximize statistical weight). These comparisons of vertical velocity profiles between the thermal plume model and LES results allow us to find the optimal value for drag term coefficient:~$b=1 \times 10^{-4}$. Because the vertical velocity profile also depends on~$\epsilon$, these comparisons had to be done in an iterative way. Although our parameterization slightly overestimates~$\Gamma$ in the upper part of the plume, it accurately describes the external forces acting on the plume throughout most of the PBL vertical extent.

\subsection{Formulation adopted for entrainment and detrainment}
\label{subsec:entform}

\citet{Nord:94} suggests that entrainment can be seen as the mechanism that compensates the reduction in fractional coverage of a non-detraining thermal due to its acceleration:

\begin{equation}
\epsilon = \frac{\beta_1}{w_u} \frac{\partial{w_u}}{\partial z}
\end{equation}

where~$\beta_1 \in [0,1]$ is a parameter introduced by~\citet{Rio:10} to account for real-case conditions which may depart from the ideal theoretical situation. Using equations~\ref{eq:verticalw} and~\ref{eq:GammaRio}, this leads to:

\begin{equation}
\epsilon = \frac{\beta_1}{1 + \beta_1} \left( a \, \frac{B}{w_u^2} - b \right)
\label{eq:epsappr}
\end{equation}

This formulation was compared to our LES results for Mars. We found~$\epsilon$ tends to be underestimated in the main part of the plume (mixed layer), and overestimated in the surface layer. The alternative approach by \citet{Greg:01} (which corresponds to~$b=0$ in equation~\ref{eq:epsappr}) does not yield favorable results either: entrainment rate is acceptable only in the region of the plume above the surface layer. 

The intensity of Martian convection makes it difficult to use simple linear laws for~$\epsilon$ and to reconcile distinct entrainment modes along the vertical extent of the plume. To address this issue, we propose to use a power-law fit between~$\epsilon$ and~$\Gamma / w_u^2~$: 

\begin{equation}
\epsilon = \mathcal{E}_1 \left(\frac{\Gamma}{w_u^2}\right)^{\mathcal{E}_2} = \mathcal{E}_1 \left(a\frac{B}{w_u^2} - b \right)^{\mathcal{E}_2} .
\label{eq:epsilon}
\end{equation}

Consistent results are obtained for all LES cases listed in tables~\ref{tab:1} and~\ref{tab:2} for~$\mathcal{E}_1 = 0.037$ and~$\mathcal{E}_2 = 0.63$.

\citet{Rio:10} argue that detrainment rate can be simply parameterized as proportional to~$B/{w_u}^2$ above the inversion layer ($\delta = \mathcal{D}_1 \, B / w_u^2$ for~$B < 0$) and constant below it ($\delta = \mathcal{D}_2$ for~$B \geq 0$). Detrainment rate parameterizations have proven to be difficult to fit to detrainment computed directly from LESs. Acceptable results (though with less precision than for entrainment rates) can however be obtained by performing fits to spatially and temporally averaged LES results, which yields the values~$\mathcal{D}_1 = -0.67$ and~$\mathcal{D}_2 = 4 \times 10^{-4}$.

Non-dimensional entrainment, detrainment and mass flux are used to discretize equation~\ref{eq:conservationphi} along the vertical. Details on the mathematical formulation of the model and its discretization are included in supplementary material (section 4). A normalization flux~$F_c$ must be used to get the entrainment and detrainment mass fluxes E and D from $\epsilon$ and~$\delta$. We follow \citet{Hour:02} who proposed to compute~$F_c$ using the horizontal/vertical aspect ratio~$r$ of PBL convective cells (cf. section 4 in the supplementary material document). Martian LESs predict aspect ratios ranging from 1 to 3 depending on local time and chosen scenario. Dry terrestrial LES studies also show variations of the aspect ratio, with values ranging from 1 to 5. Given this variability, and the fact that aspect ratios on Mars are difficult to infer from existing measurements, we choose to keep the aspect ratio~$r$ as a free parameter. Adjusting~$r$ mainly changes the potential temperature profile in the surface layer for a given mass flux. Therefore,~$r$ can be set by using potential temperature profiles from LESs. For a wide range of realistic conditions, we found that~$r=1$ gives satisfying results, while any value below~$0.7$ yields insufficient mixing.

\subsection{Treatment of downdrafts}
\label{subsec:downdraft}

The downdraft term in equation~\ref{eq:turbulenceEq} is analogous to the updraft term, hence can be modeled with a similar approach. The equivalent of equation~\ref{eq:conservationphi} for a downdraft is:

\begin{equation}
\frac{1}{F_d} \, \frac{\partial F_d \, \phi_d}{\partial z} = \epsilon_d \, \phi - \delta_d \, \phi_d
\label{eq:conservation}
\end{equation}

where~$F_d$ is the downdraft mass flux, and~$\phi_d$ is the value of~$\phi$ in the downdraft.

In the parameterization for the updraft sub-column, the estimated buoyancy of the plume is used to compute a vertical velocity profile for an air parcel, accounting for drag forces. Relating downdrafts to pure buoyant motions is not as straightforward as for updrafts. Downdrafts with positive buoyancy in their initiation part are often found in Martian LESs. This is also a common problem for Earth models \citep{Couv:05}. Because of these uncertainties on underlying mechanisms, parameterizations based on vertical velocity are not considered as relevant for now. Instead, we choose to use a simpler parameterization for downdrafts. 

We found in LESs that the downdraft-to-updraft ratio of mass flux~$F_d/F_u$ is constant with height in the mixing layer (see Figure 2 in supplementary material). This ratio is~$\sim -0.8$, an absolute value less than 1, as could be expected from the behavior of a typical thermal plume whose compensating flux is downward. Furthermore, this value is, to first order, similar for all local times and all simulated cases. Hence we obtain~$F_d$ from~$F_u$ with a proportionality law in the mixed layer and a decreasing linear law in the surface layer:

\begin{equation}
F_d = - \zeta \, F_u \quad \textrm{with} \quad \zeta = \max \left( 0.8 \, , \, 4 \, \frac{z}{z_i}  + 0.6 \right)
\label{eq:fufd}
\end{equation}

Since~$F_d < F_u$, this parameterization induces a slow compensating subsidence in the environment, meaning that~$w_e \neq 0$ hence~$\Omega < 0$. Considering fractional coverages for downdrafts in Figure~\ref{fig:alphaVCSD}, we estimate~$\Omega$ to be about -3\%. This correction is negligible given the rough setting of the drag parameter~$b$ to calculate~$\Gamma$: reasonable results are also obtained with~$\Omega = 0$.

\begin{figure}
\begin{center}
\noindent\includegraphics[width=0.75\columnwidth]{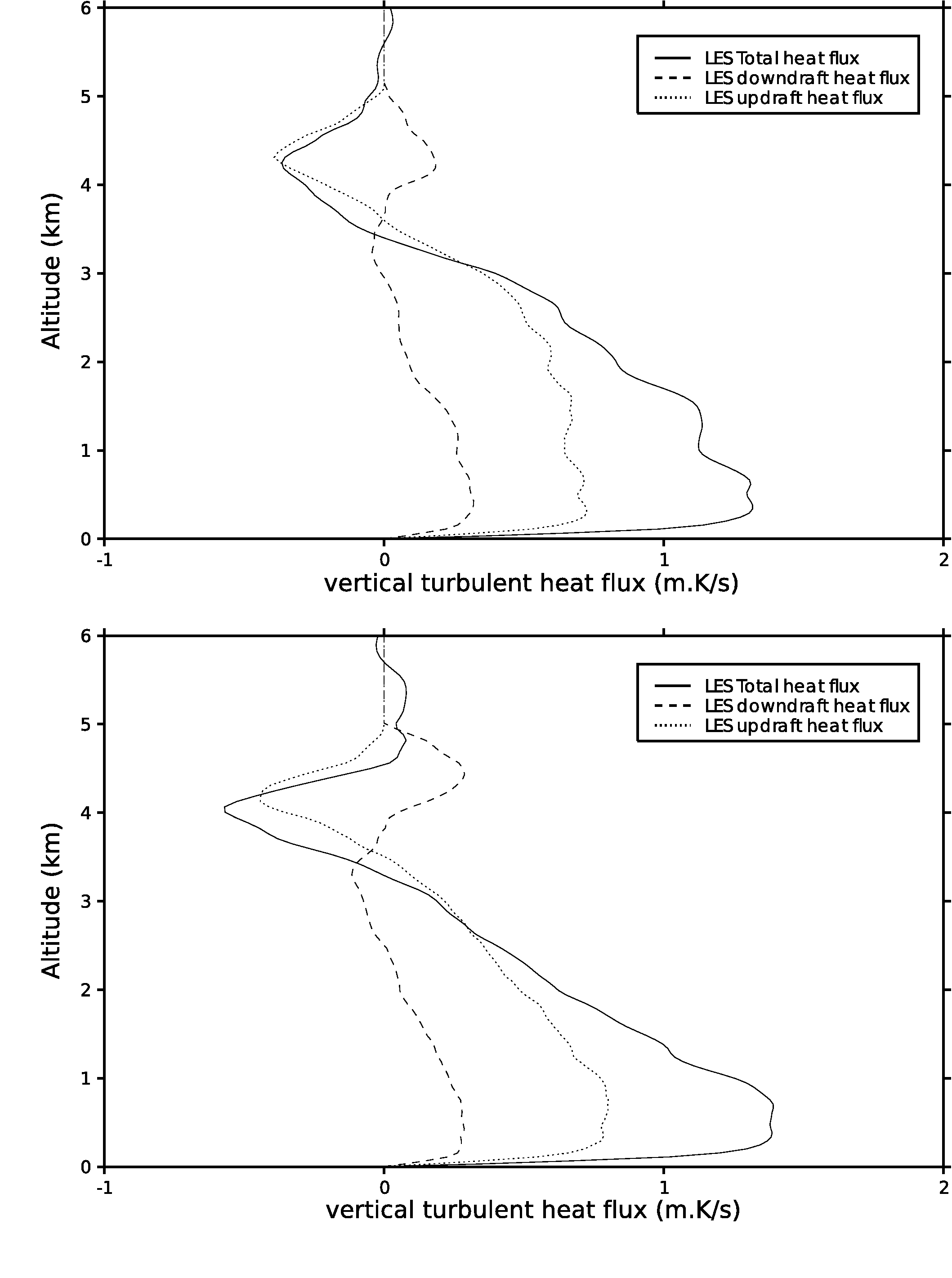}
\caption{ Contributions of downdrafts (dashed line) and updrafts (dotted line) to the total heat flux (solid line) for the LES of cases A [top] and I [bottom] at local time 13:00. Similar results are obtained for all timesteps and all cases.}
\label{fig:thfdecomp}
\end{center}
\end{figure}

To complete the prescription of downdraft heat fluxes, we simply set a potential temperature profile for downdrafts from environment properties:

\begin{equation}
\theta_d = \xi \, \theta_e \quad \textrm{with} \quad \xi = \min \left( 1 \, , \, \frac{1}{400} \frac{z}{z_i} + 0.9978 \right)
\end{equation}

where coefficients are chosen so that parameterized turbulent heat fluxes (diurnal cycle and vertical profile) are satisfyingly reproduced in the surface layer and mixed layer compared to LES results in Figure~\ref{fig:thfdecomp}. Good results are also obtained in the inversion layer provided that a fine enough vertical grid is adopted to resolve overshoots (see section \ref{sec:resultsthermals}).

Prescribing downdraft quantities allows us to compute temperature tendencies through downdraft heat flux divergence. Because entrainment and detrainment rates are unknown, it is not possible to deduce tracer and momentum transport in the downdraft, unless one computes entrainment and detrainment rates from prescribed mass flux and prescribed potential temperature profiles. The latter approach has proven to be difficult for downdrafts. As a result, for now in the thermal plume model, only downward transport of tracer and momentum related to compensatory subsidence in the environment is taken into account. Future work is needed to overcome this limitation.

\section{Surface layer parameterization}
\label{sec:sl}

Since surface-atmosphere fluxes control the amount of heat and momentum leaving the surface, and being transported and mixed by PBL eddies, parameterizations for PBL mixing must be coupled to a surface layer scheme to compute surface-atmosphere interactions. Here we present a surface layer parameterization based on a Monin-Obukhov Richardson formulation, and coupled to turbulent gustiness estimated from our Martian thermal plume model. This improved surface layer scheme is important to predict more accurately the near-surface atmospheric state measured by Mars landers (see section \ref{subsec:resultsSL}).

The principle of surface layer schemes is that momentum and heat fluxes between surface and atmosphere, denoted~$u_\star^2$ and~$u_\star \theta_\star$ respectively, are computed by multiplying the difference between surface and atmospheric values with aerodynamic conductances:~$\Lambda_m = C_D U_0$ for momentum and~$\Lambda_h = C_H U_0$ for heat (where~$U_0$ is the norm of the horizontal wind at the model first layer at altitude~$z_1$ and~$C_D$ and~$C_H$ are momentum and heat bulk transfer coefficients). One of the simplest parameterizations for surface-atmosphere momentum and heat fluxes is the neutral coefficient formulation, where~$C_D = C_{DN} = \left(\kappa / \ln(z_1/z_0) \right)^2$ and~$C_H = C_{DN}$ ($\kappa$ is Von Karman's constant and~$z_0$ the roughness length). This simple parameterization tends to overestimate exchange fluxes, and does not take into account atmospheric stability in the computation of transfer coefficients, which is especially problematic on Mars where near-surface stability is highly variable on the diurnal scale.

Using a Monin-Obukhov similarity theory to compute transfer coefficients on Mars yields more realistic results than a neutral bulk scheme \citep{Habe:93pbl,Till:94,Savi:99,Mart:09}. Section 5 in the supplementary material details the formulation of our Monin-Obukhov surface layer scheme, the computation of the bulk Richardson number~$R_i$, and our choice for the stability coefficients and functions~$f_m$,$f_h$. In this model, transfer coefficients for heat and momentum~$C_D$ and~$C_H$ are~$C_D = f_m(R_i) \, C_{DN}$ and~$C_H = f_h(R_i) \, C_{HN}$.

We can use our new thermal model to further improve this surface layer scheme, especially as far as the coupling between surface and mixed layers is concerned. Indeed, LESs show that turbulent horizontal motions associated with non-local turbulence contribute to surface-atmosphere exchange fluxes; yet surface layer schemes are used in GCMs and MMs where this gustiness is left unresolved. To remedy to this, we follow an idea described by \citet{Rede:00} who argue that the wind~$U_0$ in aerodynamic conductances~$\Lambda_m$ and~$\Lambda_h$ should be replaced by a modified wind~$U$ combining the large-scale (synoptic) wind~$U_0$ near the surface with a gustiness wind~$U_g$:

\begin{equation}
U^2 = U_0^2 + U_g^2 
\label{eq:omegawind}
\end{equation}

$U_0$ is also what we named background wind in LESs (section~\ref{sec:sampling}).~$U_g$ reflects the intensity of PBL convective winds and can be computed following different methods. For instance, \citet{Godf:91} propose to define~$U_g = \beta w_{\star}$, where~$w_{\star}$ is the free convection velocity and~$\beta$ a constant parameter. The main advantage of linking~$U_g$ and~$w_{\star}$ in a GCM or MM parameterization is that the latter quantity can be extracted from the thermal plume model.

Is this linear relationship between~$U_g$ and~$w_{\star}$ valid for Mars? One of the limitations is that~$w_\star$ has to be redefined for Mars given the strong radiative forcing in the daytime PBL \citep[see equation 12 in][for the definition of~$w_\star$ for Martian applications]{Spig:10bl}. Taking advantage of the large number of points available in LES results, we can build a good statistical estimate of gustiness wind~$U_g = \sqrt{U^2 - U_0^2}$. Resulting statistics of gustiness speeds at each timestep can then be compared to~$w_\star$. We find that~$U_g$ is not perfectly linear with~$w_\star$ (see Figure 3 in supplementary material) owing to a threshold effect for high values of vertical velocity scale. Interestingly, this is also noted in \cite{Rede:00} with a different proxy for convective activity. We therefore use a fit to a logarithmic law defined as follows:

\begin{equation}
U_g = \log \left( 1 + 0.7\,w_\star + 2.3\,w_\star ^2 \right)
\label{eq:gustinessfit}
\end{equation}

This approach takes into account large-eddy gustiness for most cases, which ensures that coupling the thermal plume and surface layer models yields similar surface-atmosphere exchange fluxes to LESs. However, the above function is not optimal in conditions of large background winds (30 m/s in the mixed layer). Proxies of the convective activity other than~$w_{\star}$ do not solve this issue. Those difficulties to model gustiness at large background wind speeds have also been noted by \citet{Fent:10} and deserve dedicated studies.

\begin{figure*}
\begin{center}
\noindent\includegraphics[width=\columnwidth]{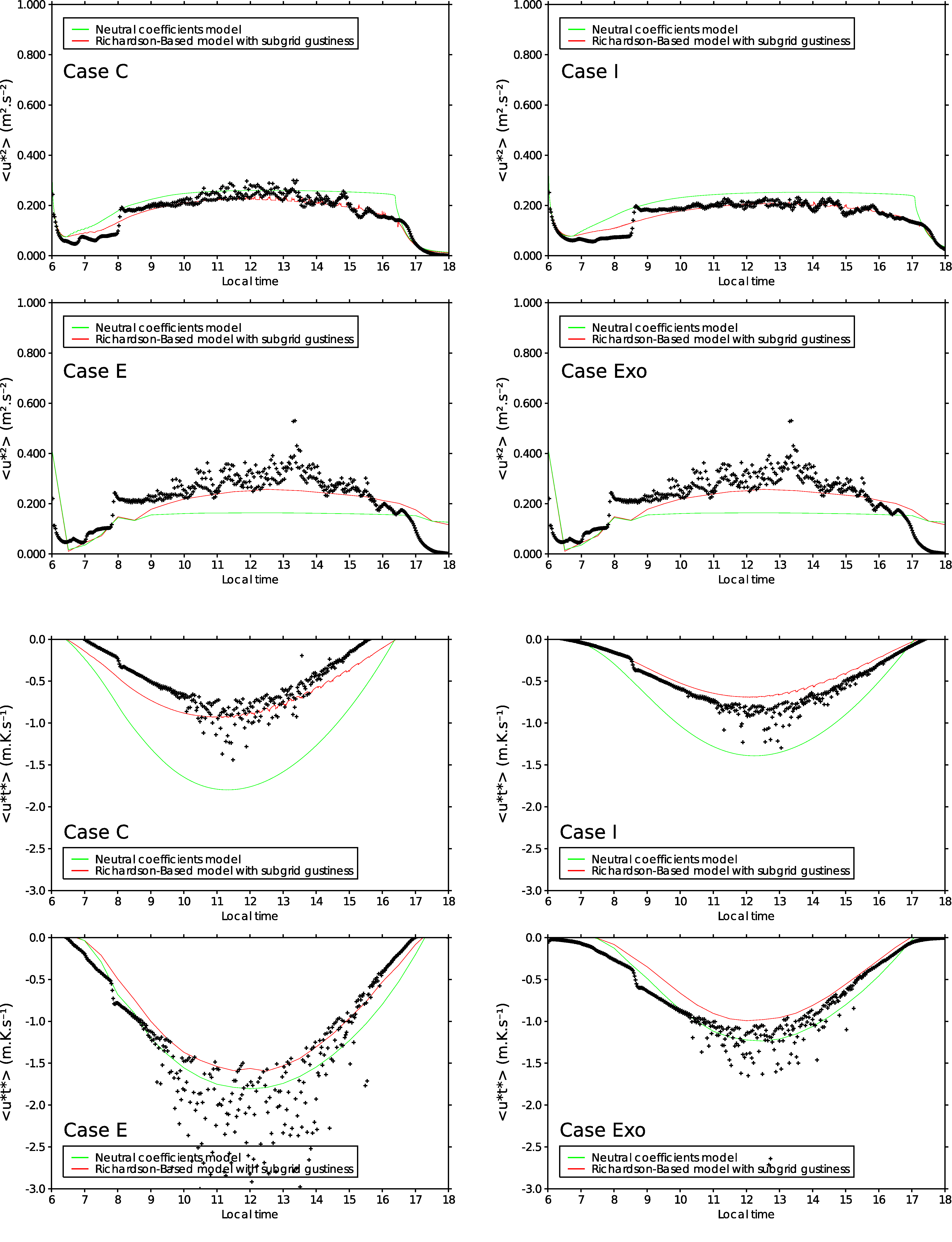}
\caption{ Comparisons of predicted~$u_\star ^2$ and~$u_\star \theta_\star$ for cases C, I, E and Exo between old and new surface layer models. Solid lines are SCM results and black crosses are LES results. SCM first level height is about 4.5m, whereas LES first model level height is about 2m.}
\label{fig:usttst}
\end{center}
\end{figure*}

Comparisons between parameterized values and LESs results for~$u_\star^2$ and~$u_\star \theta_\star$ are shown in Figure~\ref{fig:usttst}. Results comparing the neutral and the new $R_i$-based surface layer models are shown. Changes in the description of~$u_{\star}^2$ are minor yet reflect gustiness-enhanced fluxes. Changes in~$u_\star \theta_{\star}$ are important and highlight how the neutral model severely overestimates~$C_H$.

\section{Performance of the new Martian PBL schemes}
\label{sec:resultsthermals}

Here we assess the performance of single-column Martian simulations using our Martian thermal plume model and Richardson-based surface layer with turbulent gustiness. The thermal plume model only accounts for non-local turbulent transport. A Mellor and Yamada 2.5 diffusion scheme is employed for small-scale mixing, and local mixing by large eddies \citep[see][appendix B]{Hour:02}. Other physical parameterizations (namely radiative transfer) are otherwise similar to Martian LES carried out in this study (see section~\ref{sec:sampling}). Sub-surface layers are initialized with a 2-year ``warmup" run in our SCM. Model predictions are eventually validated against lander measurements in the near-surface of Mars.

\subsection{SCM without radiation}
\label{subsec:noradresults}

\begin{figure}
\begin{center}
\noindent\includegraphics[width=0.5\columnwidth]{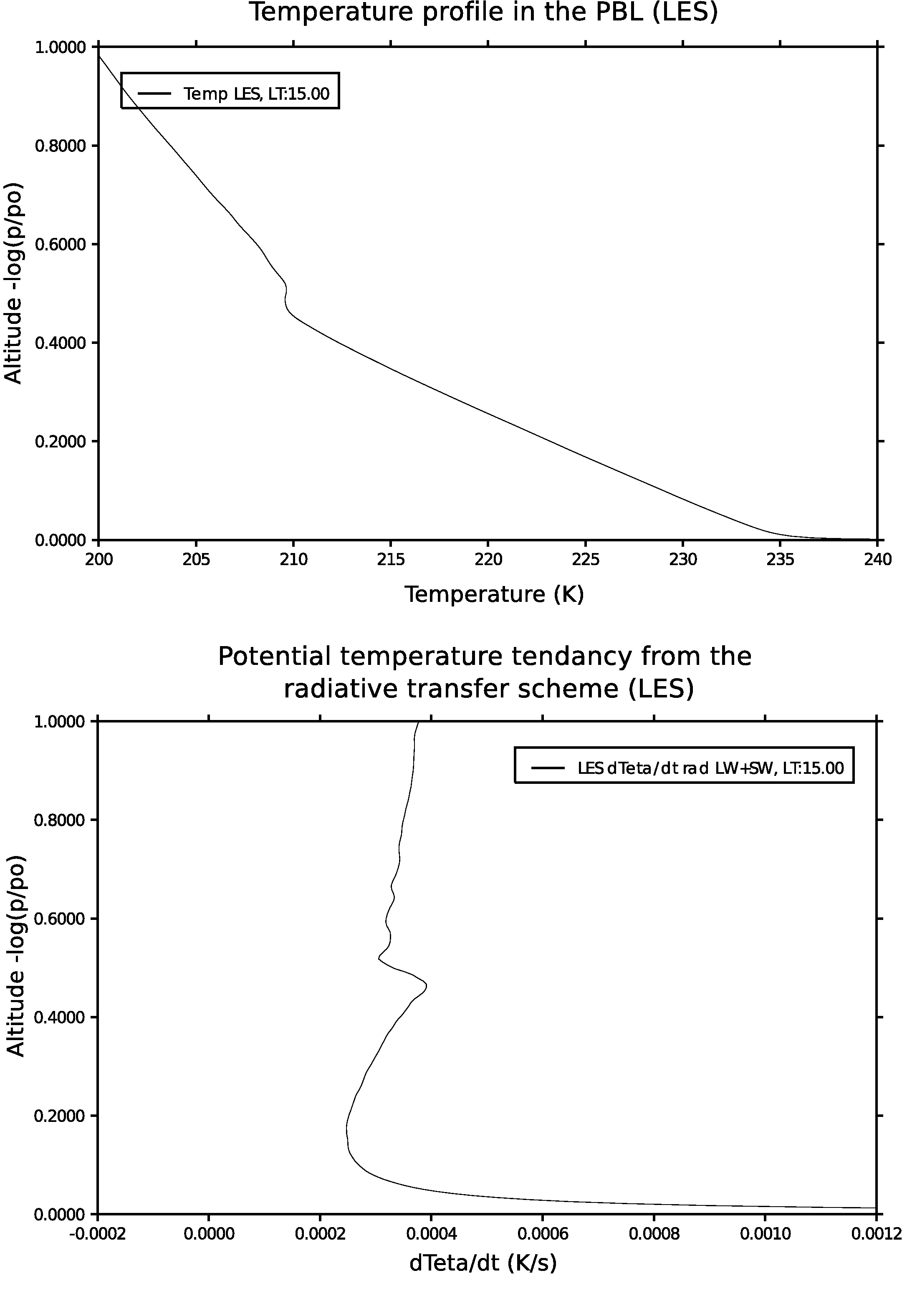}
\caption{ Slab-averaged profiles of temperature [top] and heating rates predicted by radiative transfer [bottom] at local time 15:00 for simulation case I. Heating rates include both shortwave [SW, visible wavelengths] and longwave [LW, infrared wavelengths] components. The SW contribution results mainly from direct absorption of incoming solar radiation by~$\mathrm{CO}_2$ and dust. In our case assuming well-mixed dust in the PBL, it is approximately constant with height. The variations with height of heating rates mostly arise from variations in LW heating rates \citep[.e.g.][]{Habe:93pbl}. }
\label{fig:radiation}
\end{center}
\end{figure}

Single-column simulations are first run at similar vertical resolution and integration timestep as LESs. This aims at assessing the performance of our new parameterizations alone, without any bias arising from unresolved gradients in coarse vertical discretizations. An additional specific setting is needed to pursue this aim. Figure~\ref{fig:radiation} shows that in the PBL radiation and convection are closely intertwined. Notably, the PBL top exhibits a local temperature enhancement which corresponds to the overshoot region for thermal plumes where their remaining advected heat is deposited: this causes radiative cooling at the PBL top, and warming above and below. 

\begin{figure}
\begin{center}
\noindent\includegraphics[width=0.75\columnwidth]{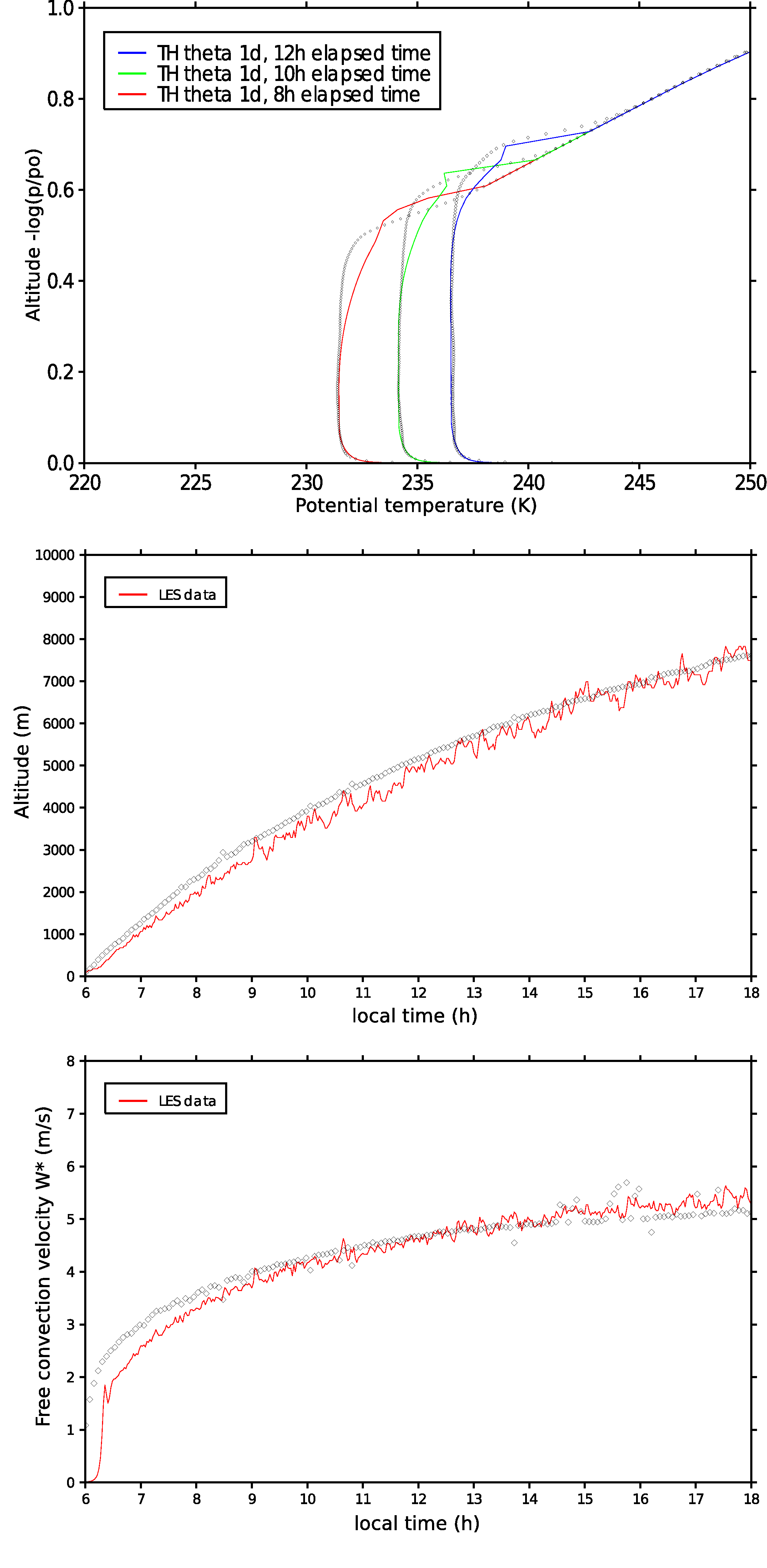}
\caption{Comparisons between high vertical resolution SCM (solid lines) and LES (diamonds). Case with radiative transfer deactivated and sensible heat flux prescribed. Potential temperature profiles after 8h,10h and 12h of simulation [top], PBL height~$z_i$ [middle] and free convection velocity~$w_{\star}$ [bottom]. Note that contributions from the local turbulence scheme are taken into account in the computation of~$w_\star$, in addition to the thermal plume model contributions.}
\label{fig:Tetanorad}
\end{center}
\end{figure}

The validity of the Martian thermal plume model, and surface layer, is first assessed without radiative transfer, in a case which resembles dry PBL convection on Earth. A fixed sensible heat flux is prescribed in~LES and~SCM to~$\overline{{w'\theta'}_0} = 1.5~$ K m/s to reproduce typical PBL dynamics obtained in case C. Potential temperature profiles in both SCM and LES are compared in Figure~\ref{fig:Tetanorad} [top]. Temperature inversion in the surface layer is well represented by the thermal plume model compared to the LES. The overshoot region near PBL top is also reproduced, although with less precision. Because this detraining layer is much more dynamic in LESs, it is inherently more challenging to parameterize precisely. However, we note that PBL top is accurately reproduced in potential temperature profiles. Mixed layer temperatures predicted by the SCM are in agreement with LESs along most of the PBL depth, despite values slightly too low in the overshoot region. PBL height in the SCM is estimated using the predicted vertical velocity profile and compared to the height at which the vertical velocity in plumes cancels out in the LES (Figure~\ref{fig:Tetanorad}, all diagnostics are domain-averaged in LES). Although a slight offset is present, PBL height is correctly predicted by the SCM, meaning that the overshoot region of thermals is adequately parameterized. Finally, we compare free convection velocity~$w_\star$ (calculated from maximum vertical eddy heat flux and PBL height according to equation 12 in \cite{Spig:10bl}) between LES and SCM in Figure~\ref{fig:Tetanorad} [bottom-right]. Predictions from the SCM in Figure~\ref{fig:Tetanorad} are found to be satisfyingly close to LES results.

\subsection{SCM with radiation}
\label{subsec:radresults}

\begin{figure*}
\begin{center}
\noindent\includegraphics[width=\columnwidth]{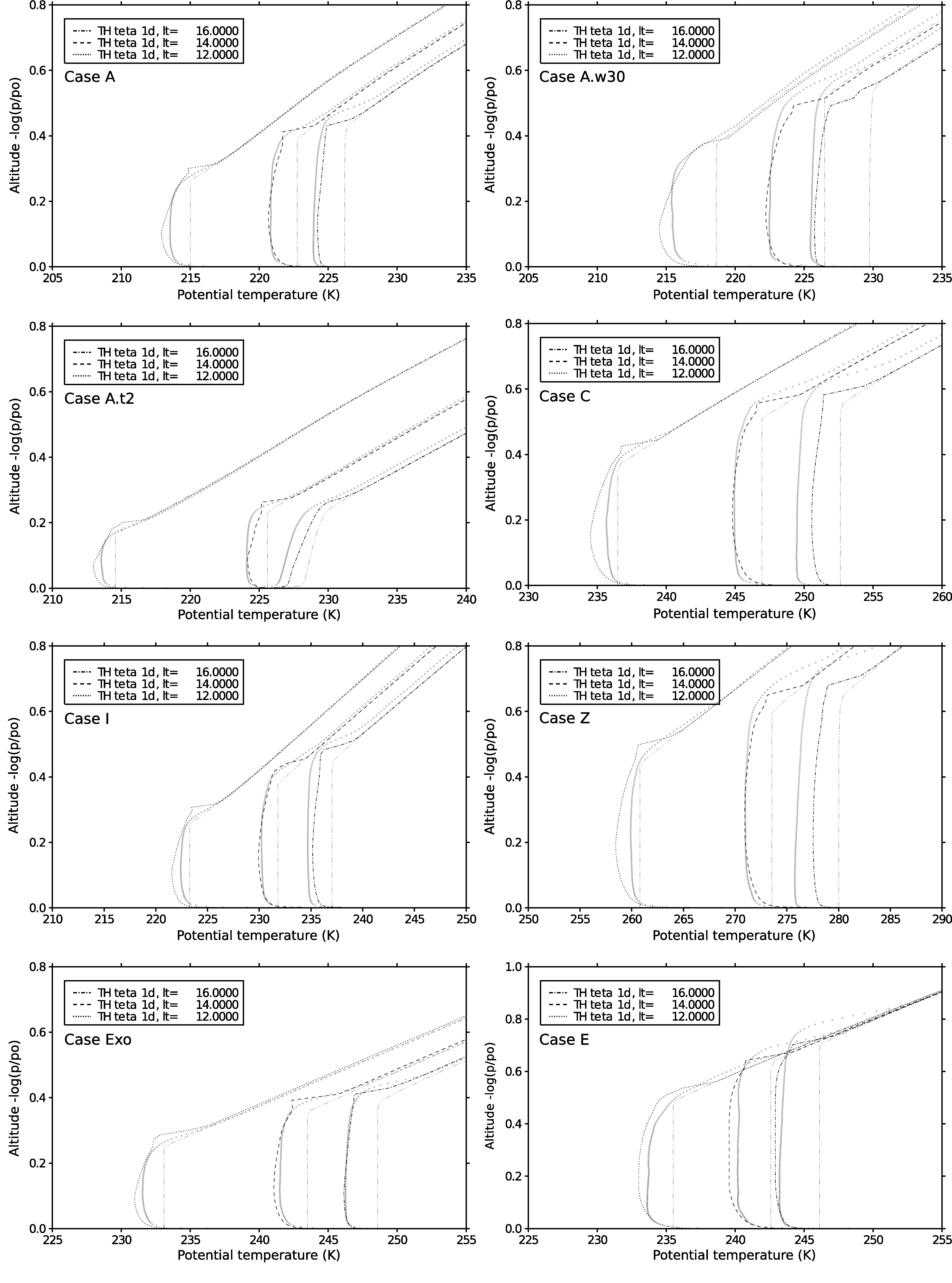}
\caption{ Potential temperature profiles from the SCM including thermals and a Richardson-based surface layer with turbulent gustiness, compared to LES results (diamonds). Results are taken at local times 12:00 (dotted lines), 14:00 (dashed lines) and 16:00 (dotted dashed lines). Results using convective adjustment scheme and old surface layer model are superimposed in triple-dotted dashed lines.}
\label{fig:resultssummary2}
\end{center}
\end{figure*}

\begin{figure}
\begin{center}
\noindent\includegraphics[width=\columnwidth]{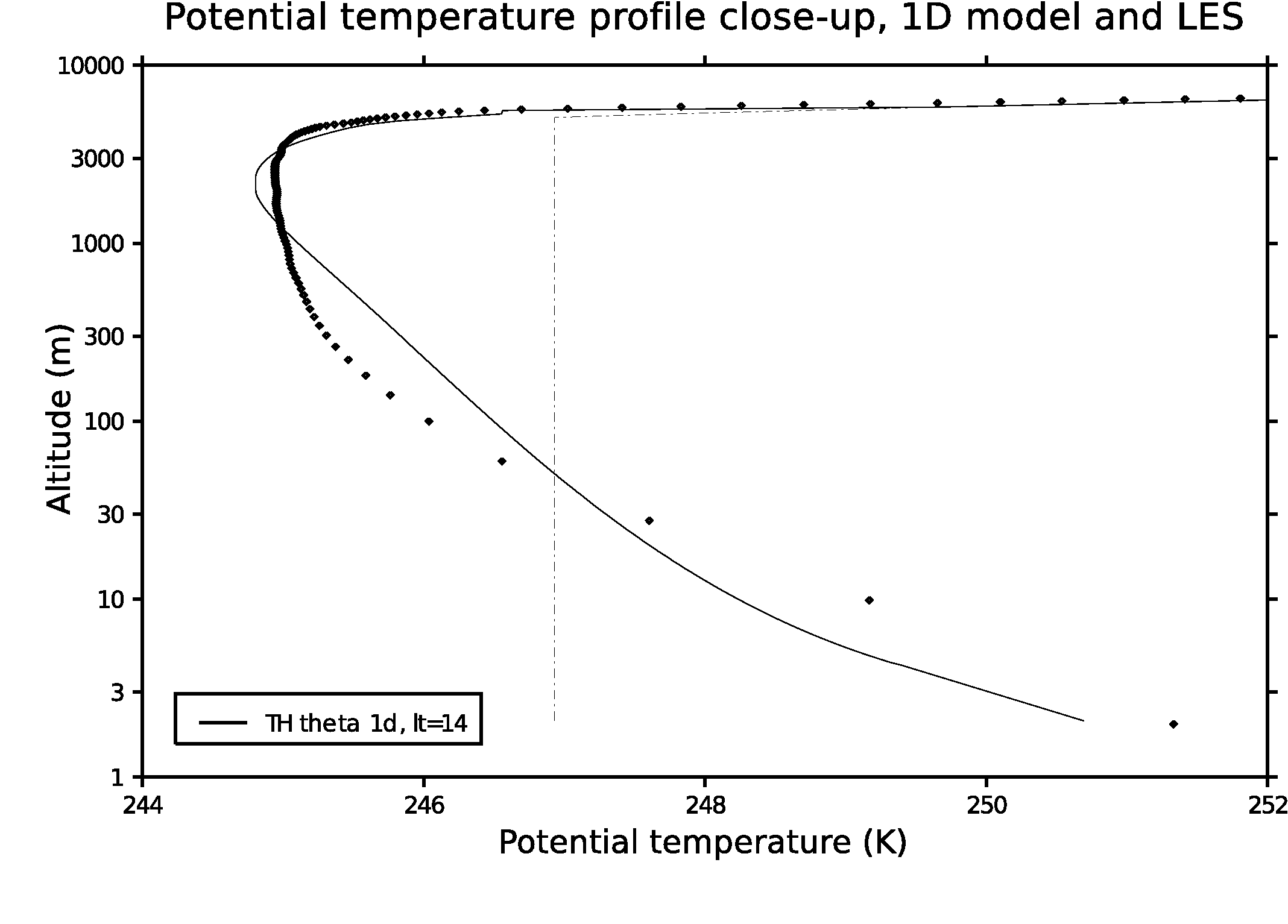}
\caption{ Potential temperature profile for case C at local time 14:00, plotted against a logarithmic altitude axis. LES results are shown as diamonds, SCM using the thermal plume model as a solid line and SCM with convective adjustment as a dotted dashed line.}
\label{fig:zoomConvadjvsTh}
\end{center}
\end{figure}

We now explore all the realistic Martian cases described in tables~\ref{tab:1} and~\ref{tab:2}. A refined vertical grid is used as in section~\ref{subsec:noradresults}. Results from the SCM are compared to LESs in Figure~\ref{fig:resultssummary2}, along with results obtained with dry convective adjustment. Our thermal plume model for Mars ensures that temperatures in the mixing layer are satisfyingly close to those predicted by LESs. The new scheme is especially efficient in maintaining a superadiabatic layer near the Martian surface (Figure~\ref{fig:zoomConvadjvsTh}), which is a key characteristic of the daytime PBL in this environment \citep{Smit:06}. This will ensure in Mars GCMs, MMs, and SCMs, that surface-atmosphere exchange fluxes and near-surface temperature profiles will be correctly estimated (especially in very convectively active cases). The prediction of the inversion layer is also satisfactory, as well as the reproduction of daytime growth of PBL height. Late afternoon temperatures in the SCM tend to be too warm, not only in the PBL but also in the troposphere.However, in this part of the convective PBL, SCM results are closer to LES case C.large than LES case C. These LESs differ by a slightly lower model top and a much larger domain extension. About 30 thermal plumes are found at a given time in case C.large, hence improving the statistics. Furthermore, convective cells are less constrained by the boundary conditions and can be represented in their full horizontal extent. Velocity scales~$w_\star$ in the SCM are in agreement with LES results (see Figure 4 in the supplementary material), which indicates that PBL convective activity is well reproduced by a SCM using our Martian thermal plume model. 

\begin{figure*}
\begin{center}
\noindent\includegraphics[width=\columnwidth]{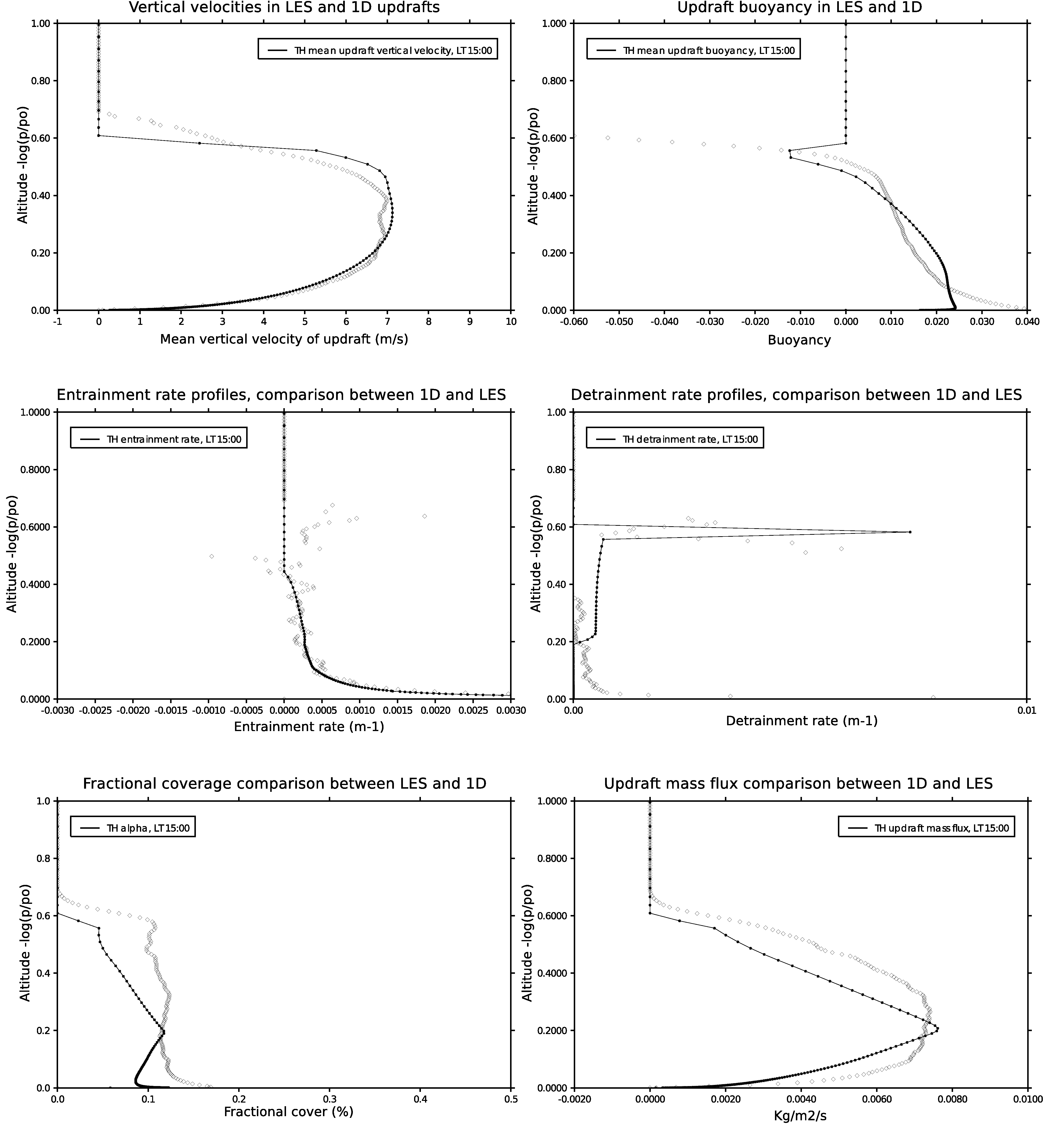}
\caption{ Intermediate variables of the thermal plume model for case C at local time 15:00. From top to bottom, left to right: vertical velocity, updraft buoyancy, entrainment rate, detrainment rate, fraction coverage, updraft mass flux. In all plots, LES results are shown as diamonds and SCM results as solid lines with bullets.}
\label{fig:internal}
\end{center}
\end{figure*}

Key ``intermediate" variables in the thermal plume model are the vertical velocity and buoyancy which determine entrainment and detrainment rates. In turn, those variables being coupled, it is difficult to get correct buoyancy and vertical velocity without correct entrainment and detrainment rates. Fractional coverages and total mass fluxes are also key diagnostics to check the robustness of our thermal plume model. Figure~\ref{fig:internal} shows that those parameterized intermediate variables in the thermal plume model compare well with predicted variables in LES integrations. This demonstrates that our thermal plume model is capable of parameterizing the non-local transport through PBL convective structures in a physically-consistent way. While being overall satisfyingly accounted for by our thermal plume model, mass-flux (and, consequently, fractional coverage) profiles tend to exhibit a peaking shape, which reflects the assumptions made on the source profile in the surface layer (see section 4.1 in the supplementary material). An alternate formulation for this source term, perhaps more suitable for the radiatively-controlled lower Martian PBL, would be needed to improve these diagnostics in the thermal plume model.

\subsection{SCM with radiation and coarse vertical grid (``GCM-like")}
\label{subsec:gcmlike}

\begin{figure}
\begin{center}
\noindent\includegraphics[width=0.75\columnwidth]{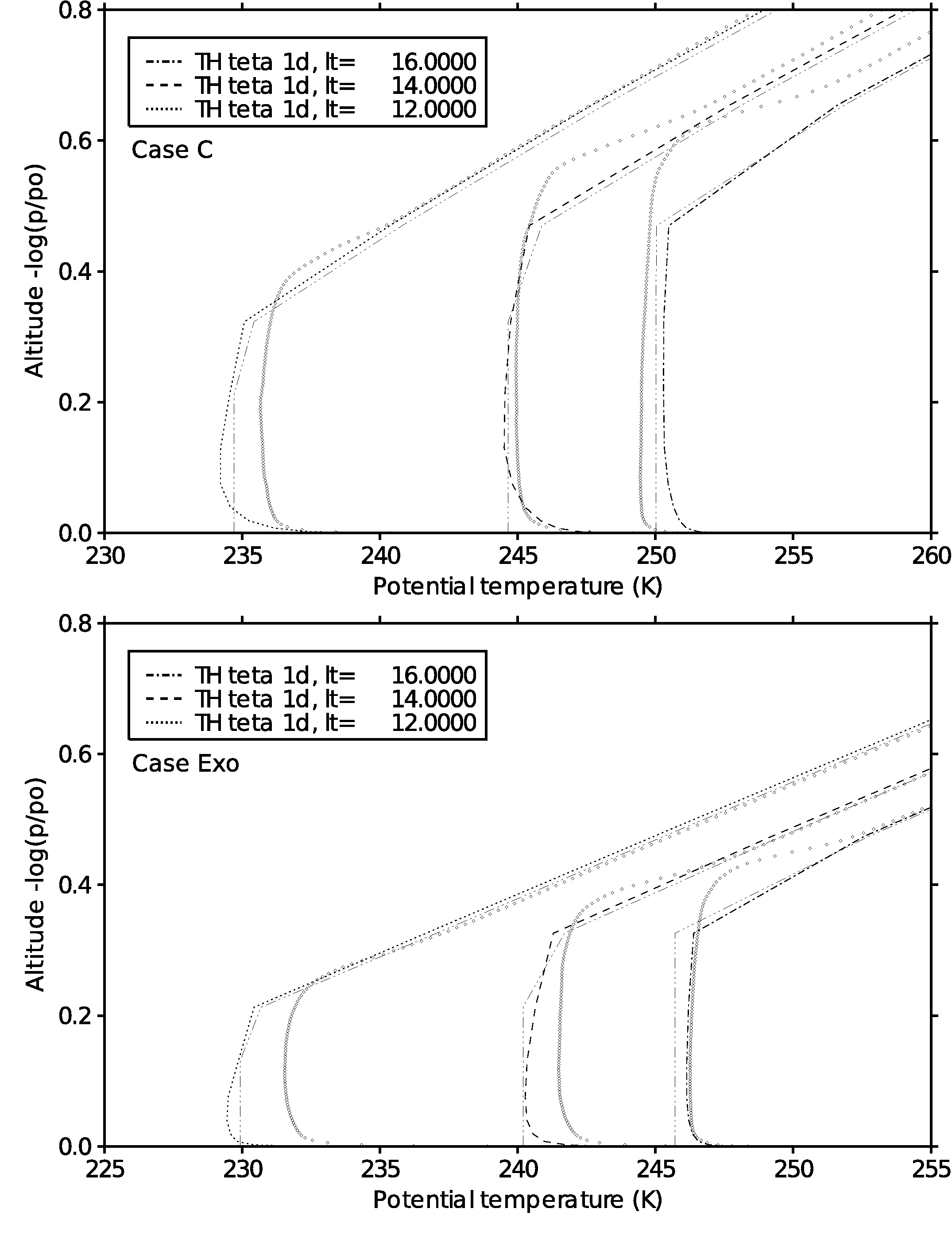}
\caption{ Temperature profiles from the SCM using the Mars LMD GCM vertical and temporal resolution, including the thermal plume model and a Richardson-based surface layer at different local times (dotted, dashed and dotted dashed lines), compared to LES results (diamonds). Results using the convective adjustment scheme and neutral coefficients surface layer model are superimposed in triple-dotted dashed lines.}
\label{fig:resolution}
\end{center}
\end{figure}

\begin{figure*}
\begin{center}
\noindent\includegraphics[width=\columnwidth]{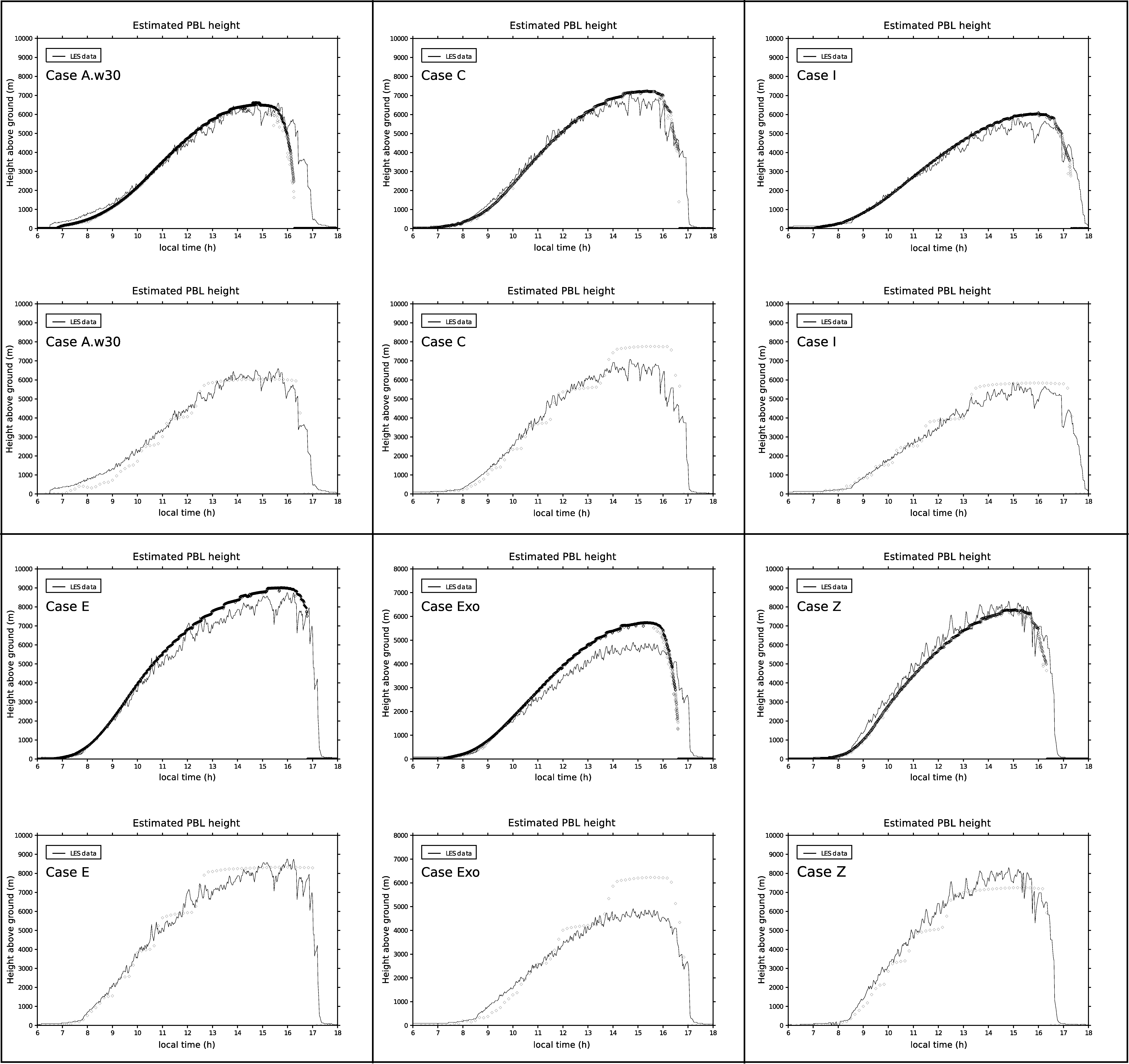}
\caption{ Comparisons of maximum height reached in the thermals between LES (solid line) and SCM (diamonds). Vertical scales can be different between figures. For a given case, top figures are obtained using high resolution and bottom figures are the corresponding GCM-like resolution results.}
\label{fig:zmax32}
\end{center}
\end{figure*}

To work most adequately, the thermal plume model should use a vertical grid with fine enough resolution to resolve gradients of intermediate quantities in the model (e.g buoyancy profiles), especially in the superadiabatic region above the surface and the overshoot region in the vicinity of the PBL top. It is difficult to achieve such vertical resolution in Mars GCMs which often use about one level for each kilometer in the PBL (and refined resolution near the surface). We thus test the thermal plume model with a SCM adopting typical GCM vertical grids (e.g. for Mars LMD GCM, 33 levels up to 130 km) and physical timesteps (4 per Martian hour). The vertical levels within the PBL in this configuration are set approximately to~[2~m, 15~m, 60~m, 160~m, 360~m, 0.7~km, 1.2~km, 2~km, 3.2~km, 4.7~km, 6.6~km, 8.9~km, 11.6~km] above the surface. Some adaptations are needed for our local diffusion scheme to yield reliable results (see section 6 in the supplementary material). We evaluate the accuracy of the thermal plume model for PBL temperature profiles in Figure~\ref{fig:resolution}, PBL height in Figure~\ref{fig:zmax32}, convective velocity scale~$w_\star$, and vertical eddy heat flux (Figure 4 and 5 respectively in the supplementary material). Although slightly less accurate than results obtained with a finer vertical grid in section~\ref{subsec:radresults} (especially in the overshoot region near the PBL top), the thermal plume model used on a coarse, ``GCM-like", vertical grid enables reliable estimates of PBL-related quantities resolved in LESs. The superadiabatic layer near the Martian surface is well reproduced by the thermal plume model compared to LES results. The good agreement shown in Figure~\ref{fig:zmax32} between the SCM and LES predictions of PBL height is also a particularly satisfying point. LES predictions of PBL height for most cases in table~\ref{tab:1} are supported by radio-occultation measurements \citep{Hins:08,Spig:10bl}. Furthermore, PBL height using our Martian thermal plume model are consistent with LES predictions, contrary to existing PBL parameterizations for Mars which tend to underestimate PBL height compared to LES \citep{Tyle:08}. 

\subsection{Comparisons with lander measurements}
\label{subsec:resultsSL}

To further validate our Martian thermal plume model and surface layer scheme, we compare predictions using our new schemes with lander measurements on Mars (data from Phoenix lander and Opportunity rover).

\begin{figure}
\begin{center}
\noindent\includegraphics[width=0.5\columnwidth]{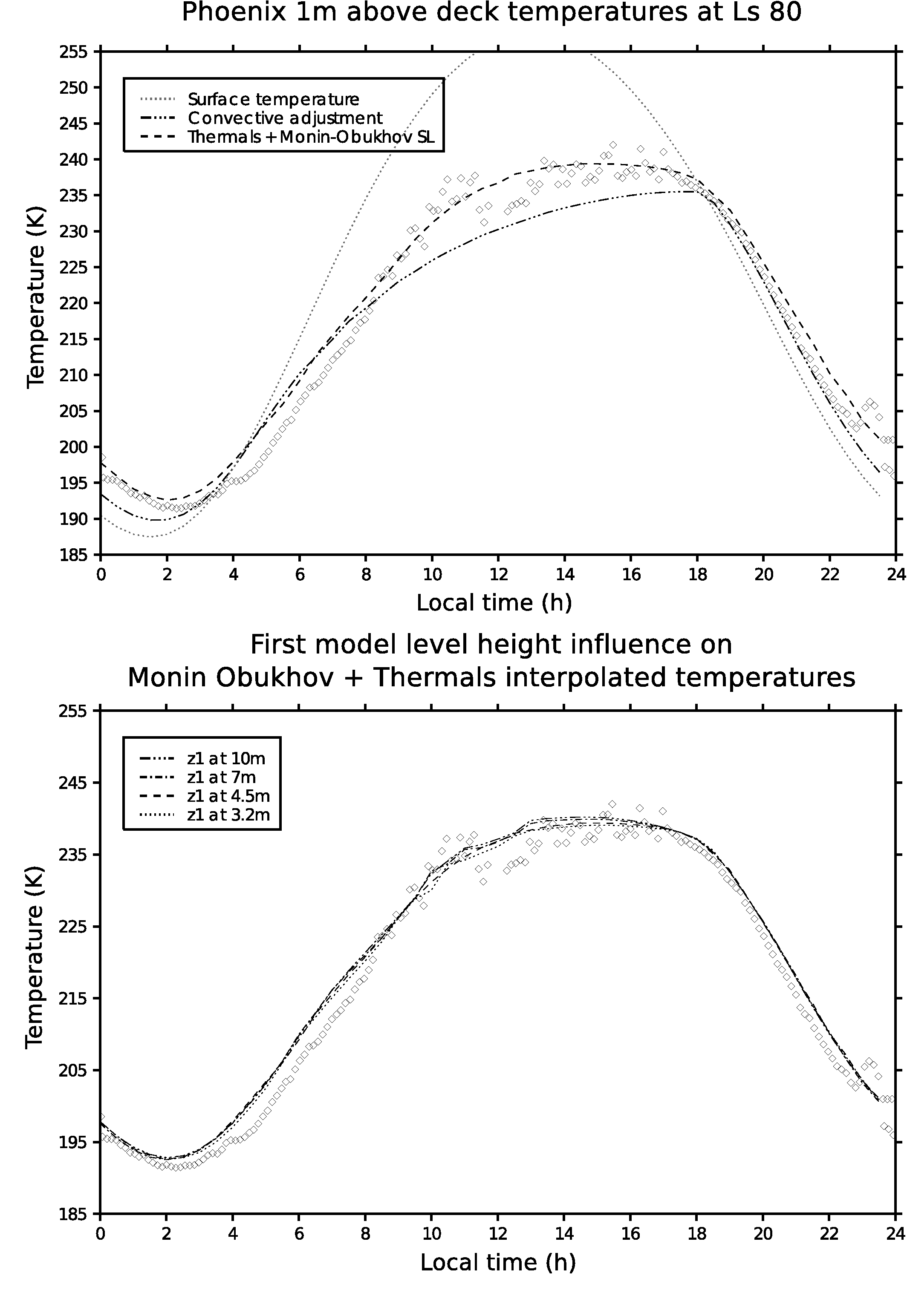}
\caption{ [Top] Phoenix measured temperatures at 1m above deck and SCM results. Phoenix data is plotted in diamonds and results from the SCM with lines. Convective adjustment temperature at 1m (dotted dashed line) is obtained by running the SCM with convective adjustment using a first model level at 2m (rover deck itself is at 1m above ground). Results using the thermals and improved surface layer (dashed line) are obtained by running the SCM with a first model level at 4.5m and running a Monin-Obukhov interpolation at 2m. [Bottom] Phoenix measured temperatures at 1m above deck and Monin-Obukhov interpolation with varying first model level height~$z_1$ between 3.2m and 10m. Phoenix data is plotted in black diamonds and results from the SCM in black lines (see legend for linestyles). (Since convective adjustment predicts sharp temperature transitions between surface and atmosphere, a Monin-Obukhov interpolation is not possible with this PBL scheme. Because the idea behind convective adjustment is to produce a neutral potential temperature gradient in the PBL, we run our SCM including convective adjustment with a first model level at 2m and take the temperature value at this level for the comparison with data.)}
\label{fig:PhoeTemp}
\end{center}
\end{figure}

Data at~$L_s=80^{\circ}$ from the temperature sensor of the Phoenix polar lander \citep{Davy:10} are compared with results from our SCM. Surface pressure is extracted from the MCD ($p_s=848$ Pa), surface roughness from \cite{Hebr:12} ($z_0 = 0.27$~cm), and dust opacity from Thermal Emission Spectrometer [TES] observations ($\tau = 0.18$). Albedo is also extracted from TES measurements~$\mathcal{A}=0.24$; a value of thermal inertia~$\mathcal{T}=165$ tiu is used instead of the TES value of nighttime apparent thermal inertia ($\mathcal{T}=250$ tiu) to better reflect observed nighttime temperatures (and the overall shape of the diurnal cycle). A constant background wind of~$20$~m~s$^{-1}$ is prescribed in the free atmosphere. Figure~\ref{fig:PhoeTemp} [top] shows a comparison between the 2m temperature sensor, and SCM results, consistently interpolated using Monin-Obukhov similarity theory from the first model level at 4.5~m above ground to 2~m above ground. Predictions with our new PBL schemes are satisfyingly close to measurements, especially compared to the use of convective adjustment which underestimates afternoon temperature by at least~$5$~K. Figure~\ref{fig:PhoeTemp} [bottom] shows the results obtained through varying the height of the first layer in the SCM from 3m to 10m. Parametrization is robust to changes in vertical discretization, meaning that fluxes and gradients are correctly represented. 

\begin{figure*}
\begin{center}
\noindent\includegraphics[width=\columnwidth]{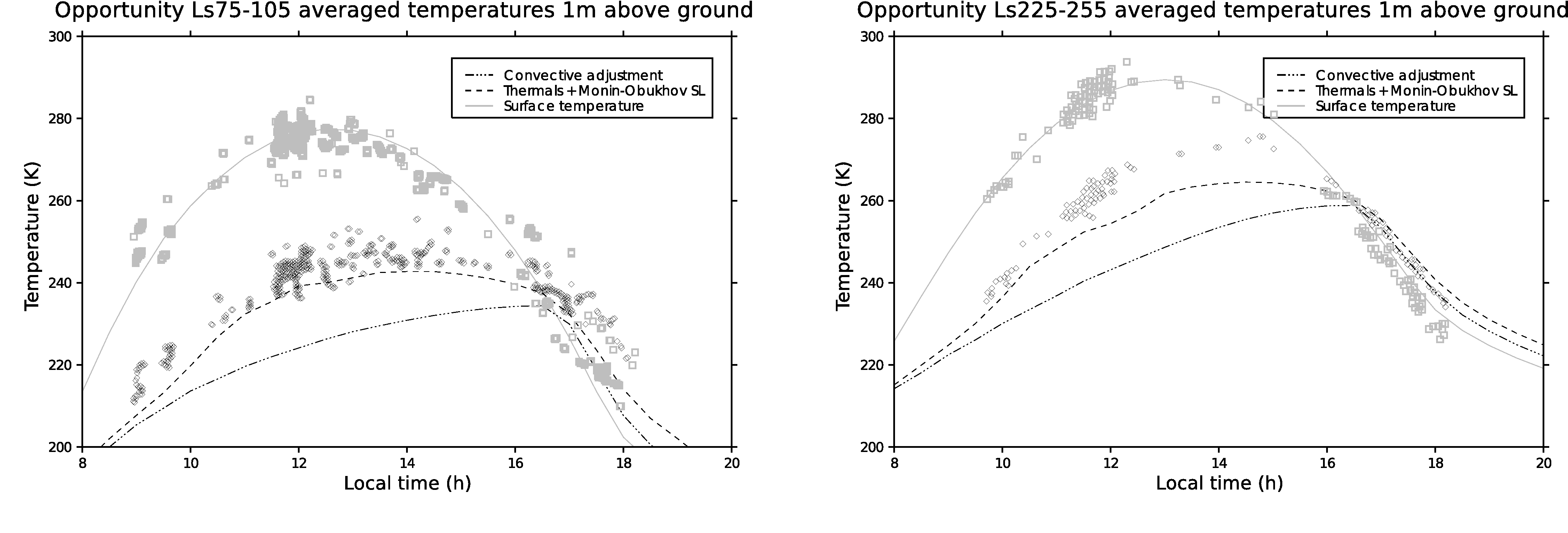}
\caption{ Opportunity measured temperatures at 1m above ground and surface temperatures averaged between~$L_s = 75^{\circ}$ and~$L_s = 105^{\circ}$ (left) and between~$L_s = 225^{\circ}$ and~$L_s = 255^{\circ}$ (right), from downward-looking mini-TES. Rover data for surface temperature is shown as gray squares and temperatures at 1m as black diamonds. SCM results are shown as lines. Convective adjustment temperature at 1m is obtained by running the SCM with convective adjustment using a first model level at 1m (triple-dotted black line). Results using the thermals and improved surface layer are obtained by running the SCM with a first model level at 4.5m and running a Monin-Obukhov interpolation at 1m (dashed line). SCM surface temperature is similar for both configurations and is shown with a plain gray line. Data is from \citet{Smit:06}.}
\label{fig:OppTemp}
\end{center}
\end{figure*}

We then choose two series of data acquired by miniTES 1m above ground \citep{Smit:06} on board the Opportunity rover close to the Martian equator: one in northern summer ($L_s = 75-105^{\circ}$) and one in northern winter ($L_s = 225-255^{\circ}$). SCM settings are defined from similar sources as for Phoenix ($z_0 = 0.92~$ cm, $p_s = 634$~Pa and~$\tau = 0.27$ in summer, $p_s = 679$~Pa and~$\tau = 0.72$ in winter). Thermal inertia~$\mathcal{T}=120$~tiu and albedo~$\mathcal{A}=0.14$ for summer, and \{$\mathcal{A}=0.23$;~$\mathcal{T}= 290$~tiu\} for winter, are obtained by fitting the predicted diurnal cycle of surface temperature with measurements (TES values are~$\mathcal{T}=280$~tiu and~$\mathcal{A}=0.18$). Distinct locations for the rover at the two considered seasons is a likely explanation for this difference in ground properties. Figure~\ref{fig:OppTemp} shows results for the two seasons. Results using the thermal plume model and Monin-Obukhov surface layer are in agreement with the data. Predicted values are too cold by a few kelvins, but results with our new model offer a significant improvement compared to the use of convective adjustment which leads to a severe underestimation (as much as~$15$~K) of daytime near-surface temperatures.

Interestingly, model predictions appear closer to observations in northern summer at low dust loading than in northern winter when the atmosphere is dustier. More measurements are clearly needed to address this question. However, a possible cause is that the Monin-Obukhov approach is less valid when dust opacity is higher. The validity of the Monin-Obukhov similarity theory on Mars is actually still left to be confirmed. The problem stems from the assumption of constant vertical heat flux made on the explicit form of the generic stability functions~$\phi_m$ and~$\phi_h$ (see supplementary material). This assumption might be flawed on Mars where radiative forcing near the surface is strong, especially when dust loading in the atmosphere is large. The Monin-Obukhov limitations in situations of strong radiative forcing require theoretical studies which are left as future work. 

\subsection{An example of use in a GCM}

We included our thermal plume model, and surface layer parameterization, in a version of the LMD-GCM \citep{Forg:99} with recent improvements of the radiative transfer \citep{Made:11}. The vertical discretization is the one described in section~\ref{subsec:gcmlike}. We use 4 physical timesteps per Martian hour instead of 2 in the version of \cite{Forg:99}. This timestep refinement yields more accurate results from the thermal plume model (compared to LES), especially as far as the coupling between radiation and convection is concerned, while being also beneficial to other parameterizations (e.g. cloud formation). 

Contrary to existing PBL parameterizations for the Martian atmosphere, our new thermal plume model enables us to estimate to first order, and map, key PBL variables such as convective velocity scale~$w_{\star}$ and maximum vertical eddy heat flux~$\langle w' \theta' \rangle_{\textrm{\footnotesize{max}}}$. Given those two variables, profiles of vertical eddy heat flux and vertical velocity variance can then be reconstructed from the Martian similarity relationships in \cite{Spig:10bl}. In addition to this, the maximum intensity~$w_{\textrm{\footnotesize{max}}}$ of vertical winds inside daytime PBL updrafts and downdrafts can be estimated by:

\begin{equation}
w_{\textrm{\footnotesize{max}}}^{u} \sim 2.75 \, w_{\star} \qquad w_{\textrm{\footnotesize{max}}}^{d} \sim 1.75 \, w_{\star}
\end{equation}

\begin{figure*}
\begin{center}
\noindent\includegraphics[width=\columnwidth]{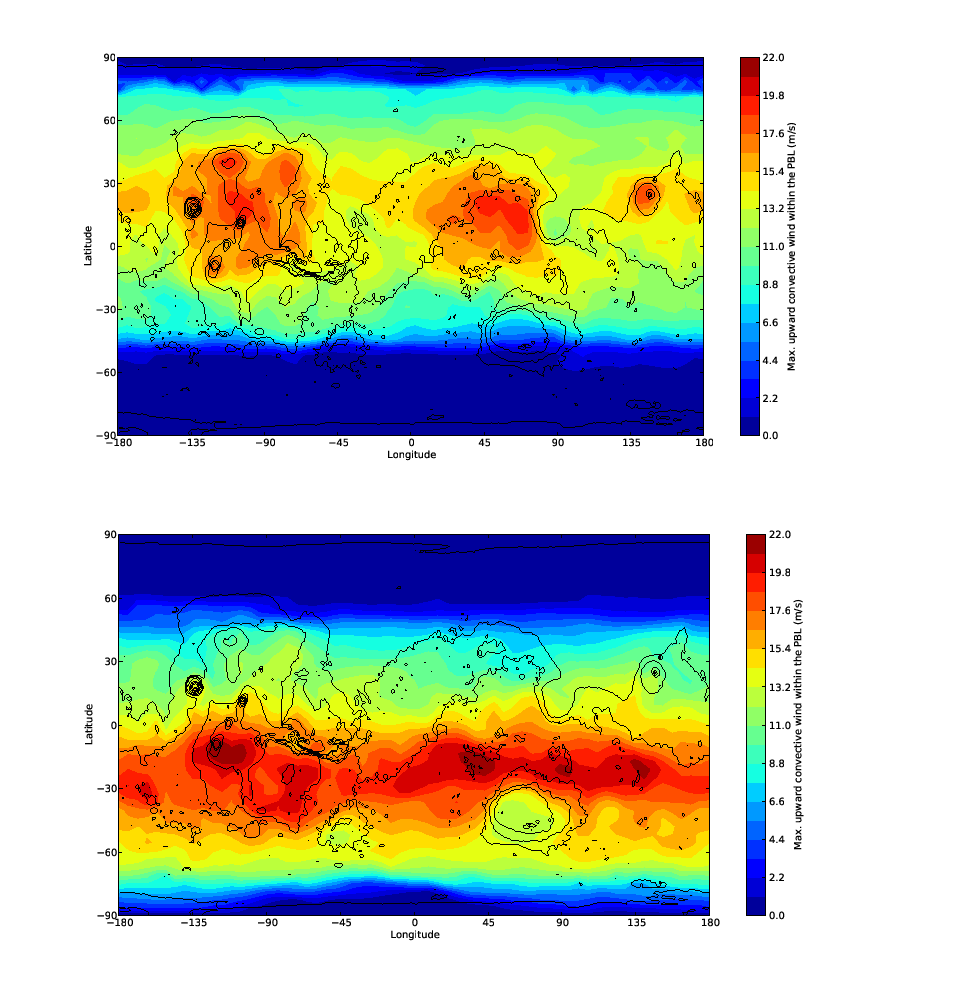}
\caption{ Longitude-latitude map of the estimated maximum velocities inside updrafts in the convective PBL at northern summer and winter solstices, obtained from the thermal plume model included in the Mars LMD GCM \citep{Forg:99,Made:11}. Results for all grid points are shown at local time 13:00. }
\label{fig:GCMmap}
\end{center}
\end{figure*}

GCM maps for~$w_{\textrm{\footnotesize{max}}}^{u}$ are given in Figure~\ref{fig:GCMmap}. The convective activity is maximum either in low-albedo (Syrtis Major) or high-topography (Tharsis, Elysium, southern high cratered terrains) areas, while it is diminished within giant impact craters (Hellas, Argyre). This is in agreement with radio-occultation measurements \citep{Hins:08} and LESs \citep{Spig:10bl}. Another possible use of~$w_{\star}$ is to use equation~\ref{eq:gustinessfit} to map near-surface horizontal gustiness due to PBL convection.

\section{Conclusion}

We propose an adaptation of the terrestrial plume model of \citet{Hour:02} and \citet{Rio:08} to the intense Martian PBL convection. Modifications are needed in the Martian case, because in this low density, radiatively-controlled environment, what is named ``shallow" convection on Earth is actually not so shallow. We show that downdrafts contribute significantly to the transport in the PBL, as they represent non-local turbulence that cannot be handled by a local diffusion scheme (e.g. Mellor and Yamada scheme). Our reformulated parameterization of thermals is inspired by terrestrial deep convection schemes using three-parcel decompositions for each grid point: an updraft, a downdraft and an environment. In addition to our new thermal plume model for Mars, we propose an improved surface layer model. Our new surface layer is based on Monin-Obukhov similarity theory, uses a Richardson formulation for stability functions, and accounts for the contribution of turbulent gustiness in surface-atmosphere fluxes. With the new thermal plume model and surface layer parameterization, the near-surface atmospheric structure predicted by climate models is much closer to the one predicted by turbulence-resolving LESs. Furthermore, GCMs and MMs are then able to compute more accurate predictions for key turbulent quantities, such as free convection velocity scale, PBL height, and maximum vertical turbulent heat flux. This permits extensive mapping of such quantities, or reconstructions of turbulent profiles, without having to run LESs. The code of our Martian thermal plume model is available upon request to the authors.

Characterizing daytime convective plumes through LESs makes our parameterization prone to the same limitations as the LES approach. Martian LESs performed thus far use periodic boundary conditions, and assume flat topography, uniform surface properties (albedo, thermal inertia) and dust loading in the domain. While the general principles of the thermal plume model presented in this paper remain valid, its ``fine tuning" could be affected by those limitations. For instance, a small-scale crater, or contrasts of albedo, might alter how frequent and strong updrafts would be. Sampling these structures will yield different constraints for the thermal plume model. Hence we expect our scheme to be refined as LESs improve in the future, and as more measurements are available to validate them.

Our new PBL parameterizations have broad implications for Martian atmospheric studies. Improvements obtained for the PBL wind and temperature structures are likely to impact the predictions of GCMs and MMs and possibly the understanding of Martian atmospheric dynamics. Moreover, how dust particles, major climatic agents on Mars, are lifted from the surface, injected in the first meters of the atmosphere, then transported above the PBL, remains an open question in which the knowledge of PBL processes is a key factor of progress. Finally, our work will hopefully allow for better estimation of atmospheric hazards met by landing spacecraft which will explore Mars in the future. Our approach for PBL parameterizations is also an extreme example which can be of interest for terrestrial meteorology. 

%%%%%%%%%%%%%%%%%%%%%%%%%%%%%%%
%%%%%%%%%%%%%%%%%%%%%%%%%%%%%%%

\section*{Acknowledgments}
We warmly thank the three reviewers whose thorough comments helped to improve and clarify the paper. We acknowledge support from CNES and ESA.

%\bibliographystyle{apalike}
%\bibliography{/home/aymeric/Work/submitted/newfred} 

\appendix
\section{List of variables and symbols}

\small
\begin{tabular}{cl}
$g$ & mean gravitational acceleration (m.s$^{-2}$) \\
$h$ & effective height (m) \\
$q$ & tracer concentration (kg.kg$^{-1}$) \\
$u$ & west-east horizontal wind component (m.s$^{-1}$) \\
$v$ & south-north horizontal wind component (m.s$^{-1}$) \\
$w$ & vertical wind component (m.s$^{-1}$) \\
$\theta$ & potential temperature (K) \\
$z_i$ & PBL height (m) \\
$\phi$ & a conserved variable \\
$\alpha$ & fractional coverage \\
$\rho$ & density (kg.m$^{-3}$) \\
$E$ & entrainment volume mass flux (kg.m$^{-3}$.s$^{-1}$) \\
$D$ & detrainment volume mass flux (kg.m$^{-3}$.s$^{-1}$) \\
$F_u$ & updraft surface mass flux (kg.m$^{-2}$.s$^{-1}$) \\
$F_d$ & downdraft surface mass flux (kg.m$^{-2}$.s$^{-1}$) \\
$F_c$ & normalization surface mass flux (kg.m$^{-2}$.s$^{-1}$) \\
$\epsilon$ & entrainment rate (m$^{-1}$) \\
$\delta$ & detrainment rate (m$^{-1}$) \\
$\epsilon^{\star}$ & non dimensional entrainment \\
$\delta^{\star}$ & non dimensional detrainment \\
$f^{\star}$ & non dimensional mass flux \\
$B$ & buoyancy force (N) \\
$\Gamma$ & drag force (N) \\
$\Omega$ & ratio of updraft to environment vertical velocities \\
$r$ & convective cell aspect ratio \\
$\Lambda_m$ & aerodynamic momentum conductance (m.s$^{-1}$) \\
$\Lambda_h$ & aerodynamic heat conductance (m.s$^{-1}$) \\
$C_D$ & momentum bulk transfer coefficient \\
$C_H$ & heat bulk transfer coefficient \\
$C_{DN}$ & momentum neutral bulk transfer coefficient \\
$C_{HN}$ & heat neutral bulk transfer coefficient \\
$U_0$ & horizontal wind norm in the first model layer (m.s$^{-1}$) \\
$U_g$ & horizontal gustiness wind (m.s$^{-1}$) \\
$u_{\star}$ & friction velocity (m.s$^{-1}$) \\
$\theta_{\star}$ & friction temperature (K) \\
$w_{\star}$ & free convection velocity (m.s$^{-1}$) \\
$\phi_m$ & Monin-Obukhov momentum stability function \\
$\phi_h$ & Monin-Obukhov heat stability function \\
$f_m$ & momentum stability function \\
$f_h$ & heat stability function \\
$P_{rt}$ & turbulent Prandtl number \\
$\kappa$ & Von Karman's constant \\
$L$ & Monin-Obukhov length (m) \\
$R_i$ & Richardson number \\
$R_{ic}$ & critical Richardson number \\
$z_0$ & momentum roughness length (m) \\
$z_{0T}$ & heat roughness length (m) \\
$R_e$ & Reynolds number \\
$R_e^{\star}$ & friction Reynolds number \\
$\nu$ & kinematic viscosity (m$^2$.s$^{-1}$) \\
\end{tabular}
\normalsize

\section{Turbulent kinetic energy (organized term)}

According to~\citet{Sieb:95}, the contribution from non-local large eddies (organized turbulence) to turbulent kinetic energy can be written as:

\begin{equation}
\mathrm{TKE}_{org} = \frac{1}{2} \, \alpha \, (1-\alpha) \left[ (u_u - u_e)^2 + (v_u - v_e)^2 + (w_u - w_e)^2 \right]
\end{equation}

where the~$u$ subscript denotes updraft and the~$e$ subscript environment. In our proposed three-column formulation, the organized term for the TKE equation is analogous although it is slightly more complex to account for downdraft quantities (denoted by a~$d$ subscript):
\begin{eqnarray}
\mathrm{TKE}_{org} = &   & \frac{1}{2} \, \alpha_u \, (u_u - \overline{u})^2 + \frac{1}{2} \, \alpha_d \, (u_d - \overline{u})^2 + \frac{1}{2} \, (1 - \alpha_u - \alpha_d) \, (u_e - \overline{u})^2  \nonumber \\
                     & + & \frac{1}{2} \, \alpha_u \, (v_u - \overline{v})^2 + \frac{1}{2} \, \alpha_d \, (v_d - \overline{v})^2 + \frac{1}{2} \, (1 - \alpha_u - \alpha_d) \, (v_e - \overline{v})^2  \nonumber \\
 		     & + & \frac{1}{2} \, \alpha_u \, (w_u - \overline{w})^2 + \frac{1}{2} \, \alpha_d \, (w_d - \overline{w})^2 + \frac{1}{2} \, (1 - \alpha_u - \alpha_d) \, (w_e - \overline{w})^2  \nonumber \\
\end{eqnarray}

\section{Computation of entrainment and detrainment fluxes from LESs results}
\label{app:subsec:entrdetr}

We define the volume mass flux entrained in the thermal plume~$E$ and the volume mass flux detrained from the thermal plume~$D$ as in \citet{Greg:01}:

\begin{equation}
\frac{\rho}{\alpha} \, \int_u{ \left[ - \frac{\partial u'\phi'}{\partial x} - \frac{\partial v'\phi'}{\partial y} \right] \, \textrm{d}x \, \textrm{d}y} = E \phi_e - D \phi_u
\label{eq:entrdetr}
\end{equation}
 
\noindent $E$ and~$D$ can be related to the updraft fractional cover~$\alpha_u$ and surface mass flux~$F_u$ through the continuity equation in the quasi-Boussinesq approximation \citep{Arak:74}:

\begin{equation}
\rho \, \frac{\partial \alpha_u}{\partial t} + \frac{\partial F_u}{\partial z} = E - D
\label{eq:massconserv}
\end{equation}

\noindent Along a vertical single column, taking into account only the vertical-related eddy covariant term of the Reynolds equation, the evolution of a conserved variable~$\phi$ is given by:

\begin{equation}
\frac{\partial \overline{\phi}}{\partial t} = -\frac{1}{\rho} \, \frac{\partial \rho \overline{w' \phi'}}{\partial z}
\label{eq:siebstart}
\end{equation}

To parameterize updraft quantities, we assume that updrafts do not directly interact with downdrafts (the vast majority of updrafts and downdrafts resolved by our LESs can be unambiguously separated, see in the main text Figure 2,  although in some rare cases immediately adjacent updrafts and downdrafts can be found). In this framework, the only exchanges occurring at the plume's boundary are between the updraft and the environment. Hence we project equation~\ref{eq:siebstart} on two fictive sub-columns only: an environment and an updraft. Following~\cite{Sieb:95}, one can then use~\ref{eq:entrdetr} and~\ref{eq:massconserv} to get separate expressions for E and D, and entrainment/detrainment rates~$\epsilon = E / F_u$ and~$\delta = D / F_u$:

\begin{eqnarray}
E(\phi_e - \phi_u) & = & F_u \frac{\partial \phi_u}{\partial z} + \frac{\partial \rho \alpha_u \overline{w'\phi'}^u}{\partial z} + \rho \alpha_u \left[ \frac{\partial \phi_u}{\partial t} - \left(\frac{\partial \overline{\phi}}{\partial t}\right)^{\mathrm{rad}}\right]  \nonumber \\
D(\phi_e - \phi_u) & = & F_u \frac{\partial \phi_e}{\partial z} - \frac{\partial \rho (1-\alpha_u) \overline{w'\phi'}^e}{\partial z} - \rho (1-\alpha_u) \left[ \frac{\partial \phi_e}{\partial t} - \left(\frac{\partial \overline{\phi}}{\partial t}\right)^{\mathrm{rad}} \right]  \nonumber \\
\end{eqnarray}

where~$\left(\frac{\partial \overline{\phi}}{\partial t}\right)^{\mathrm{rad}}$ is the tendency for the flux of~$\phi$ from radiative transfer. In LES results,~$E$ and~$D$ are computed using potential temperature~$\phi=\theta$, so that the last term in the equation is simply the heating rate computed in physical parameterizations for radiative transfer. 

\section{Technical details on the thermal plume model}
\label{app:subsec:discretization}

\subsection{Source layer}

The thermal plume model is activated whenever an unstable potential temperature profile is detected at the surface \citep{Hour:02}. Surface layer mixing cannot be represented by thermals, since small-scale turbulent eddies are at play. First levels above the surface are therefore represented by a so-called source layer, i.e. a prescribed profile of lateral entrainment a(z) in the surface layer. In these layers, the entrainment rate is given by a(z) and there is no detrainment. Different assumptions for typical profiles exist, e.g. as is proposed by \cite{Rio:08}:

\begin{equation}
 a(z) =\Upsilon \sqrt{z} \max\left(0,- \frac{\Delta \theta}{\Delta z}\right)
\end{equation}

\noindent where~$\Delta \theta$ is the potential temperature difference between two model levels,~$\Delta z$ is the associated model layer thickness and $\Upsilon$ is a normalization constant chosen so that the value of the integral of $a(z)$ over the vertical is one.

\subsection{Discretization}

Our thermal plume model is discretized along the vertical by defining non-dimensional entrainment $\epsilon^{\star}$, detrainment $\delta^{\star}$ and mass flux $f^{\star}$:

\begin{eqnarray}
\epsilon^{\star} & = & f^{\star} \Delta z \epsilon \nonumber \\
\delta^{\star}   & = & f^{\star} \Delta z \delta \nonumber \\
f^{\star}        & = & \frac{F_u}{F_c}
\end{eqnarray}

\noindent where~$\Delta z$ is the model layer thickness and $F_c$ is a normalization flux. At a given level~$k$,~$\epsilon$ and~$\delta$ are computed from~$B$ (buoyancy) and~$w^2$ using the vertical velocity equation:

\begin{equation}
\frac{1}{2}\frac{\partial w_u^2}{\partial z} = - \epsilon \, w_u^2 \, (1-\Omega)  + \Gamma
\end{equation}

\noindent Given that $f^{\star}$ is zero at ground level, from a known~$f^{\star}$ at level~$k$, one can compute the value of~$f^{\star}$ at level~$k+1$ using the continuity equation:

\begin{equation}
\epsilon - \delta = \frac{1}{F_u} \frac{\partial F_u}{\partial z}
\end{equation}

in the following discretized form:

\begin{equation}
\Delta f^{\star} = \epsilon^{\star} - \delta^{\star}
\end{equation}

At the end of the iteration process,~$F_c$ is computed using the closure equation (\ref{eq:closure}) so that~$F_u$ can be obtained as well as~$E$ and~$D$ through:

\begin{eqnarray}
E = \frac{F_c \epsilon^{\star}}{\Delta z} \nonumber \\
D = \frac{F_c \epsilon^{\star}}{\Delta z}
\end{eqnarray}

Following \cite{Hour:02}, the normalization flux~$F_c$ is the mean total surface mass flux that would correspond to a non-detraining updraft whose entrainment profile is the source profile:

\begin{equation}
F_c=\frac{w_{\mathrm{max}}}{r\,z_{\mathrm{max}} \sum \frac{a_k^2}{\rho_k \Delta z_k}}
\label{eq:closure}
\end{equation}

where~$w_{\mathrm{max}}$ is the maximum vertical velocity in the updraft,~$z_{\mathrm{max}}~$ is the updraft height,~$r$ is the horizontal/vertical aspect ratio of the convective cell equal to $L/z_{\mathrm{max}}$ and the sum runs from the surface up to the top of the source layer.~$\epsilon$ and~$\delta$ are multiplied by~$F_c$ to get the entrainment and detrainment volume mass fluxes E and D.

Using $E$, $D$ and $F_u$ one can integrate the conservation equation described in the main text:

\begin{equation}
\frac{\partial F_u \phi_u}{\partial z} = E \phi - D \phi_u
\end{equation}

to get $\phi_u$, hence the updraft contribution to the tendency of $\phi$, using both equation for~$\overline{w'\phi'}$ in the main text and equation~\ref{eq:siebstart} in this document.

\section{Richardson-based Monin-Obukhov surface layer with turbulent gustiness}
\label{app:sec:risl}

Computations of surface-atmosphere fluxes rely on expressions depending on bulk properties which take the form:

\begin{equation}
F_A = C_A \, (A_0 - A_{\mathrm{surf}})U_0
\label{eq:surflux}
\end{equation}

where~$F_A$ is the surface-atmosphere flux of quantity A,~$C_A$ is the corresponding bulk transfer coefficient,~$A_0$ is the value for A at the first model level,~$A_{\mathrm{surf}}$ is the surface value of A and~$U_0$ is the large-scale (synoptic) wind at the first model level. Surface-layer parameterizations are used to compute the aerodynamic conductances:

\begin{equation}
\Lambda_A = C_A \, U_0
\label{eq:aerodyncond}
\end{equation}

We detail in the main text how $F_A$ is enhanced in the GCM and MM by interpolating quadratically the synoptic wind at first level with a parameterized gustiness wind. Hence in equation \ref{eq:aerodyncond} we use the modified wind~$U$ combining the large-scale (synoptic) wind~$U_0$ near the surface with a gustiness wind~$U_g$, instead of simply~$U_0$. What is left to be described is the other term in~$\Lambda_A$ in equation \ref{eq:aerodyncond}, i.e. the prediction of bulk transfer coefficients~$C_A$. Most formulations for~$\Lambda_A$ are either functions of heat transfer coefficient~$C_H$ or momentum transfer coefficient~$C_D$. 

The determination of~$C_H$ and~$C_D$ depends on the assumed profiles for heat and momentum in the surface-layer. \citet{Moni:54} found that according to the~$\Pi$ theorem and assuming constant vertical heat flux in the surface layer, the dimensionless parameter~$\frac{z}{L}$ defines the scaling structure of the surface layer via the flux-gradient relationship for mean horizontal velocity and potential temperature:

\begin{equation}
\frac{\partial u}{\partial z} = \frac{u_{\star}}{\kappa z} \phi_m(\xi)
\end{equation}

\begin{equation}
\frac{\partial \theta}{\partial z} = \frac{\theta_{\star}}{P_{\mathrm{rt}}\kappa z} \phi_h(\xi)
\end{equation}

where~$u_{\star}$ and~$\theta_{\star}$ are friction velocity and friction temperature,~$P_{\mathrm{rt}}$ is the turbulent Prandtl number,~$\phi_m$ and~$\phi_h$ are momentum and heat universal functions and~$\xi = z/L$ (where L is the Monin-Obukhov length). Universal functions take different forms between stable and unstable atmospheric conditions. One of the most widely-used formulation for~$\phi$ functions in terrestrial models is:

\begin{equation}
\phi_m(\xi) = \left\{ \begin{array}{rl}
1+\beta_m \xi &\mbox{ if~$\xi>0$} \\
  (1-b_m \xi)^{a_m} &\mbox{ if~$\xi<0$}
       \end{array} \right.
\label{eq:stab1}
\end{equation}

\begin{equation}
\phi_h(\xi) = \left\{ \begin{array}{rl}
\alpha_{\theta}+\beta_h \xi &\mbox{ if~$\xi>0$} \\
  \alpha_{\theta}(1-b_h \xi)^{a_h} &\mbox{ if~$\xi<0$}
       \end{array} \right.
\label{eq:stab2}
\end{equation}

Different sets of parameters exist for ($\beta_m, \beta_h, b_m, b_h, \alpha_{\theta}, \kappa$) based on terrestrial field observations. We use parameters proposed by \citet{Dyer:74} for Earth, also adopted for Mars by \citet{Till:94}:  

\begin{equation}
\{\beta_m, \beta_h, a_m, a_h, b_m, b_h, \alpha_{\theta}, \kappa\} = \{5, 5, -1/4, -1/2, 16, 16, 1, 0.41\}
\end{equation}

This formulation for stability functions is not self-consistent, as~$\xi$ depends on the Monin-Obukhov length which depends on stability functions which in turn depends on~$\xi$. Although this open-form formulation can be solved by the use of iterative processes, closed-form formulations that incorporate stability dependance directly into exchange coefficients exist. \citet{Engl:95} derived an expression for stability functions~$f_m$ and~$f_h$ which varies with the bulk Richardson number~$R_i$ in the first layer:

\begin{equation}
f_m(R_i) = \phi_m^{-2}(\xi)
\end{equation}

\begin{equation}
f_h(R_i) = \phi_h^{-1}(\xi) \phi_m^{-1}(\xi)
\end{equation}

\citet{Engl:95} showed that for a stable atmosphere with~$P_{\mathrm{rt}} = 1$, stability functions can be written:

\begin{equation}
f_h(R_i) = f_m(R_i) = \left\{ \begin{array}{rl}
\left(\frac{R_{ic} - R_i}{R_{ic}}\right)^2 &\mbox{ if~$0<R_i<R_{ic}$} \\
 0 &\mbox{ if~$R_i>R_{ic}$}
       \end{array} \right.
\end{equation}

where~$R_{ic}$ is the critical Richardson number, defined as~$R_{ic} = 1/\beta_m$.
For an unstable atmosphere, i.e.~$R_i<0$ and for the chosen set of parameters, the stability functions and turbulent Prandtl number are:

\begin{eqnarray}
f_m(R_i) = \sqrt{1-16 R_i} \nonumber \\
f_h(R_i) = (1-16 R_i)^{3/4} \nonumber \\
P_{\mathrm{rt}}(R_i) = (1-16 R_i)^{-1/4}
\label{eq:ublfunctions}
\end{eqnarray}

where for the chosen set of parameters, equations (\ref{eq:ublfunctions}) are valid for all negative values of~$R_i$. \citet{Engl:95} also argue that the bulk Richardson number should be dependent on surface roughness length~$z_0$. Taking a step further, the Richardson number can also be written as a function of both the momentum and thermal roughness length~$z_0$ and~$z_{0T}$:

\begin{equation}
R_i = \frac{g}{\theta_0} h \frac{(\theta_1 - \theta_0)}{U_0^2}
\end{equation}

where~$\theta_0$ is the surface temperature,~$\theta_1$ potential temperature in the first model level and $h$ is an effective height defined as:

\begin{equation}
h =  \sqrt{z_0 z_1} \frac{(\ln{z_1/z_0})^2}{\ln{z_1/z_{0T}}}
\end{equation}

LES results show a significant difference between Richardson numbers computed from slab-averaged values and slab-averaged Richardson numbers. This difference is another illustration of the role played by turbulent statistics resolved in LESs on the prediction of averaged values used at GCM and MM scales. Following the same approach as for~$\Lambda_A$, we replace~$U_0$ with~$U$ in the computation of the bulk Richardson number. This approach also improves performance in weakly unstable conditions sometimes encountered during daytime at the poles of Mars, where~$U_0$ is small and~$\theta_1 - \theta_0$ is negative. 

The final form of the transfer coefficients in the new formulation reads:

\begin{equation}
C_D = \left(\frac{u_{\star}}{U_0}\right)^2 = f_m(R_i) \, C_{DN} = \frac{f_m(R_i) \kappa^2}{(\ln \frac{z_1}{z_0})^2}
\end{equation}

\begin{equation}
C_H = \frac{u_{\star} \theta_{\star}}{U_0 (\theta_0-\theta_1)} = f_h(R_i) C_{HN} = \frac{f_h(R_i) \kappa^2}{\ln \frac{z_1}{z_0}\ln \frac{z_1}{z_{0T}}}
\end{equation}

where the thermal roughness length~$z_{0T}$ is computed from the roughness Reynolds number~$R_{e}^{\star}$ and the Prandtl number according to \citet{Brut:82}, following a formalism also applied to Mars by \citet{Till:94}:

\begin{equation}
z_{0T} = z_0 \exp( -7.3 \, \kappa \, {R_{e}^{\star}}^{1/4} \, P_{\mathrm{rt}}^{1/2})
\end{equation}

\begin{equation}
R_{e}^{\star} = \frac{u_{\star} z_0}{\nu}
\end{equation}

where the kinematic viscosity~$\nu$ is set to~$10^{-3}$~m$^2$~s$^{-1}$ (a typical value for the Martian atmosphere).~$R_i$ is initialized with a value for~$z_{0T}$ assumed to be one tenth of~$z_0$.~$z_{0T}$ is then computed and the process is iterated until the value for~$z_{0T}$ converges toward a limit. Finally, the aerodynamic conductances for momentum and heat read:

\begin{eqnarray}
\Lambda_m = C_D U =  f_m(R_i) C_{DN}\sqrt{U_0^2 + U_g ^2} \nonumber \\
\Lambda_h = C_H U =  f_h(R_i) C_{HN}\sqrt{U_0^2 + U_g ^2}
\end{eqnarray}

where~$C_{DN}$ and~$C_{HN}$ are the neutral transfer coefficients defined in the main text.

\section{Notes on local diffusion scheme in the Martian LMD GCM}

Numerical problems are encountered with the Mellor and Yamada scheme described in \citet{Hour:02} (appendix B) when using this scheme in Martian conditions with GCM-like integration timesteps and vertical discretization. Instabilities are sometimes found in daytime in mixing coefficients, which in turn trigger strong oscillations of temperature and momentum near the surface. Oscillations are also observed in strongly stratified Martian nighttime conditions. These two issues are addressed with distinct solutions. A temporal splitting of the integration of turbulent kinetic energy solves the daytime oscillations, while nighttime oscillations are dealt with by computing a minimum mixing coefficient depending on a computed PBL height for stable conditions, following~\citet{Bros:78} and~\citet{Holt:93}.

\begin{figure}
		\centerline{\includegraphics[width = 0.75\columnwidth]{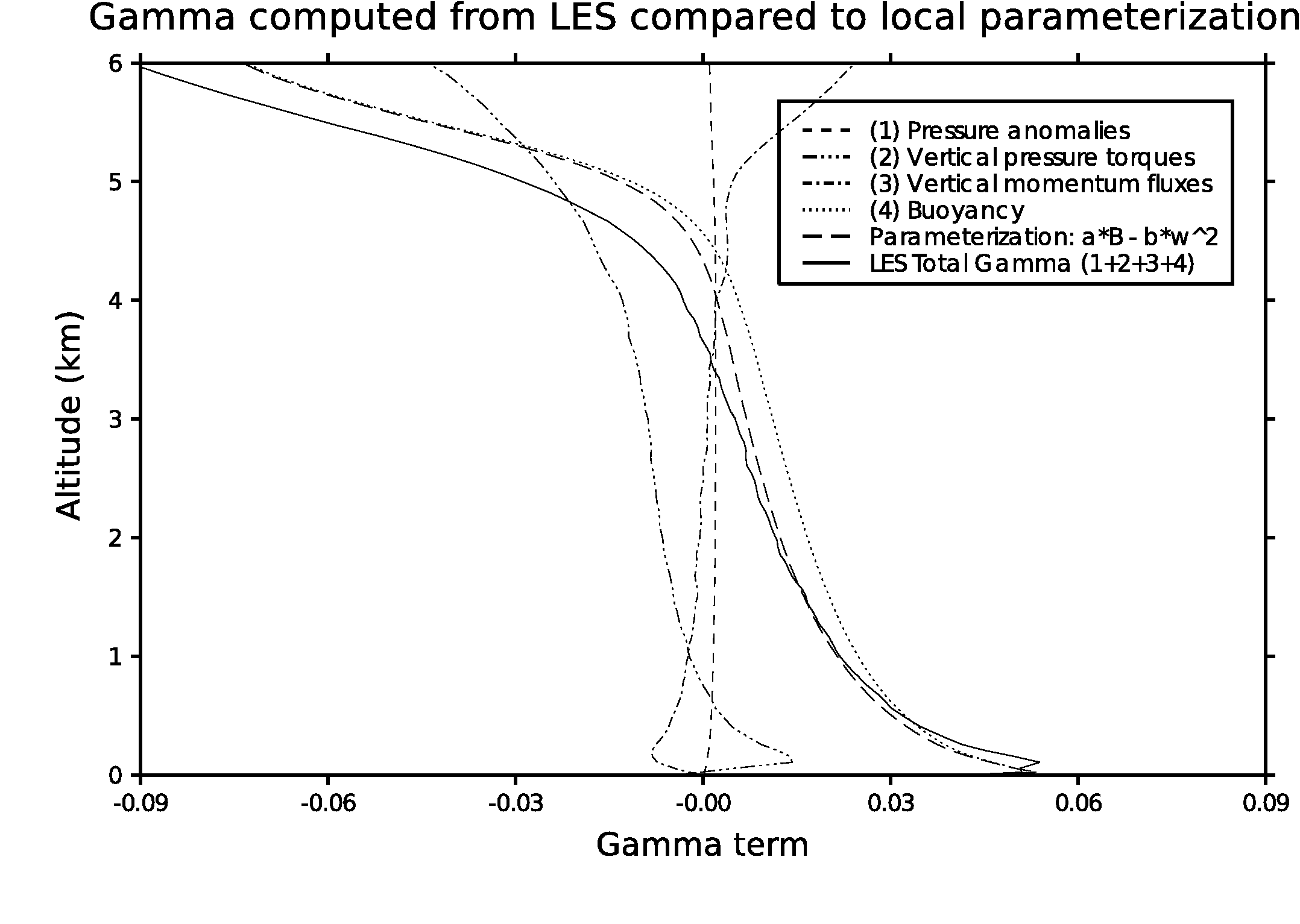}}
		\caption{ Decomposition of the terms in LES results contributing to external forces~$\Gamma$. (1), (2), (3) and (4) designate different contributions to Gamma, (4) being the buoyancy. The total of these contributions (solid line) is compared to the parameterization (long-dashed line) with the above values for coefficient~$a$ and~$b$, where we have used LES values for B and~$w^2$. LES results are from case C.large so that smooth statistics are obtained. Results are taken at local time 13:00. The correspondence between the parameterization and the LES is best below the inversion layer. Similar conclusions are drawn for other local times and simulation cases.}
		\label{fig:Gammaprofiles}
\end{figure}

\begin{figure}
		\centerline{\includegraphics[width = 0.75\columnwidth]{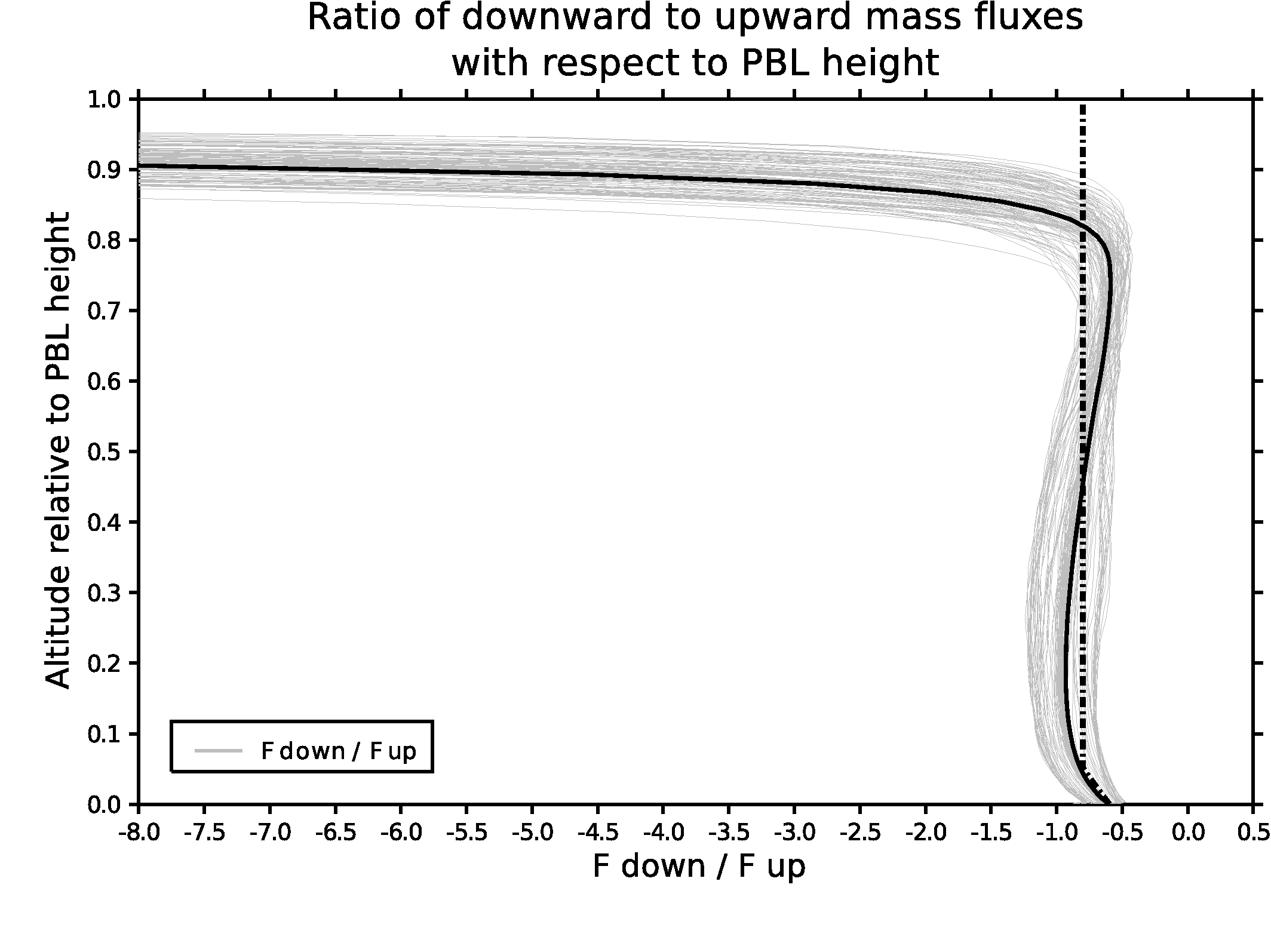}}
		\caption{ Ratios of downdraft to updraft mass fluxes as a function of altitude relative to the PBL top, for LES case C.large. Ratios are plotted for local times between 12:30 and 16:00, every 2:30 minutes (gray lines). The downward tracer being emitted from 12:00, ratios for local times before 12:30 are not computed. The average ratio is displayed as a black line. The chosen parameterization for~$f_u/f_d$ is shown as a dashed black line.}
		\label{fig:fufd}
\end{figure}

\begin{figure}
		\centerline{\includegraphics[width = 1.3\columnwidth,angle=90]{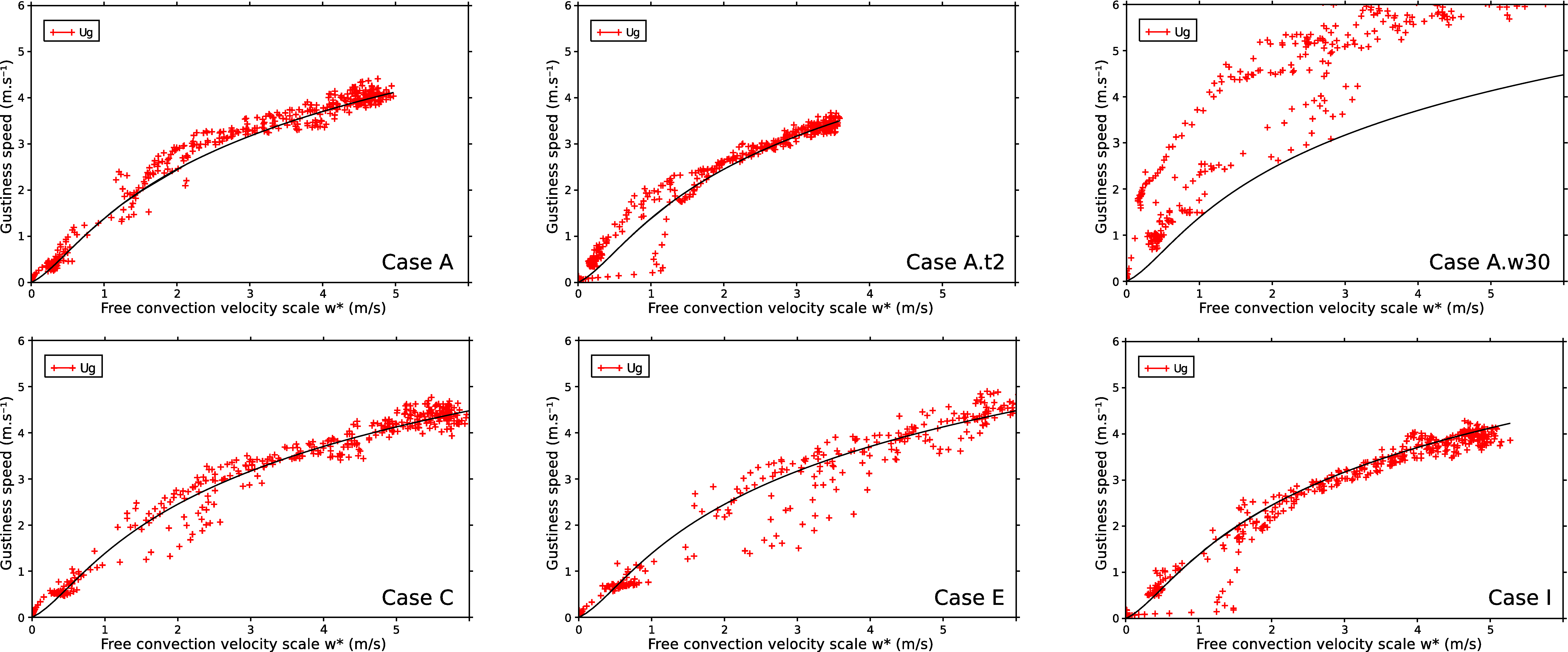}}
		\caption{ Scaling the gustiness speed in LES with free convection velocity scale. The fit of $w_{\star}$ with $U_g$ is shown as a black solid line and corresponds to $U_g = \log(1 + 0.7 w_\star + 2.3 w_\star ^2)$. LES statistics is shown as red crosses.}
		\label{fig:ugwstar}
\end{figure}

\begin{figure}
		\centerline{\includegraphics[width = \columnwidth]{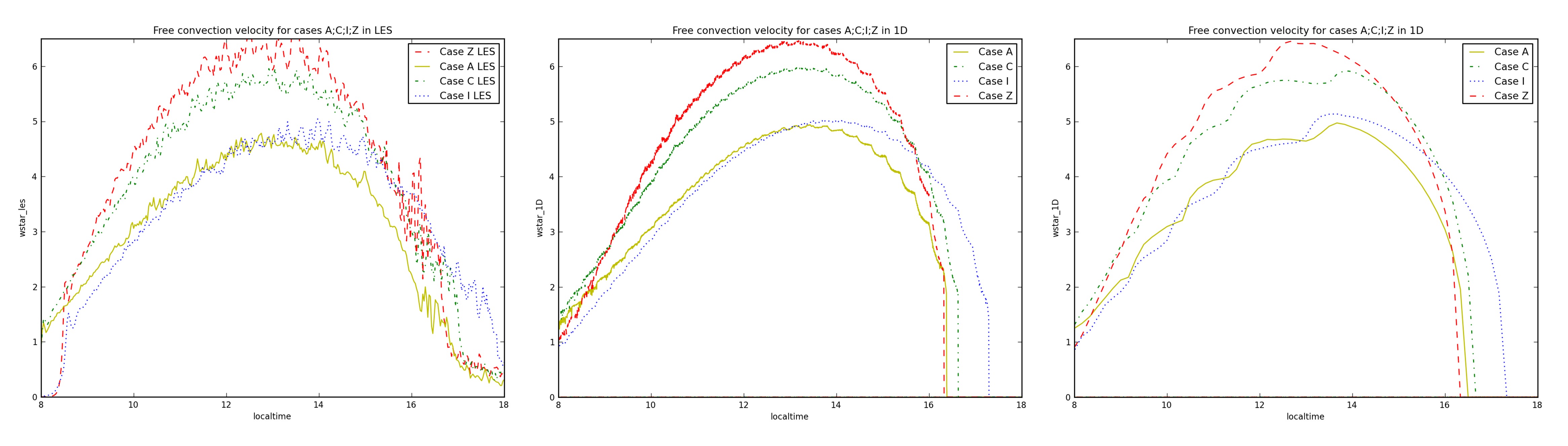}}
		\caption{ Free convection velocity scale $w_{\star}$ in LES [left] and single-column models [center and right], for the same cases as in \citet{Spig:10bl}, i.e. cases A (plain yellow), C (short-dashed green), I (dotted blue) and Z (long-dashed red). [Center] is obtained with single-column modeling using 222 vertical levels and [right] with 33 vertical levels. The high-frequency distortion observed in~$w_\star$ time series arises from the incrementation of radiative transfer tendencies every 40 timesteps to mimic what is done in GCMs; it disappears when incrementing the radiative transfer more frequently e.g. at each timestep.}
		\label{fig:wstar}
\end{figure}

\begin{figure}
\begin{center}
		\centerline{\includegraphics[width= 1.05\columnwidth]{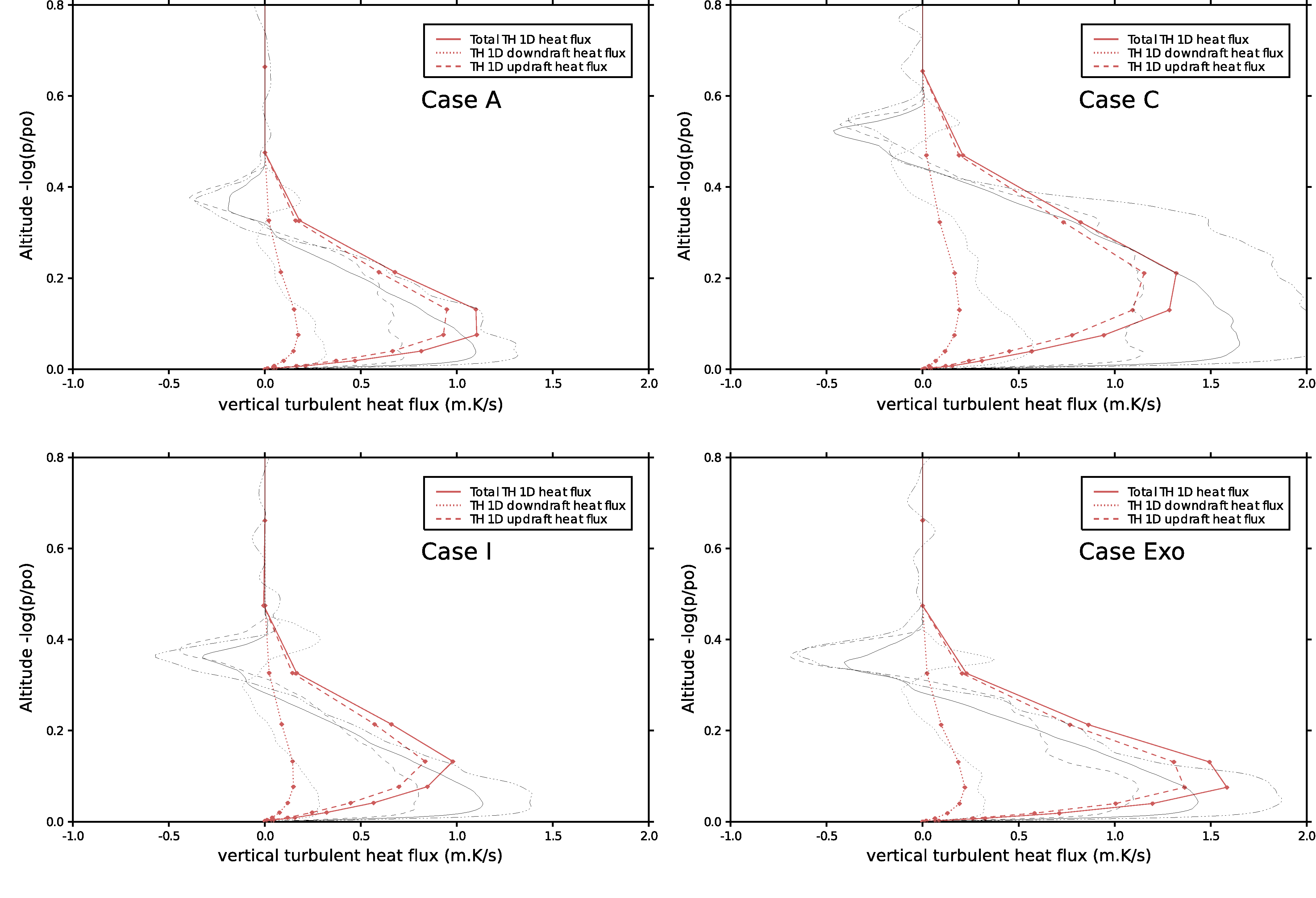}}
		\caption{ Comparison between parameterized heat flux for updrafts (red dashed line with bullets) and downdrafts (red dotted line with bullets) in the SCM (33 vertical levels are used) compared to resolved updrafts (black dashed line) and downdrafts (black dotted line) in the LES. The addition of these fluxes is shown as a red solid line and compared to the total vertical turbulent heat flux from organized structures resolved in the LES (solid line). Bullets represent single-column vertical model levels. Total heat flux resolved in the LES is shown as a black triple-dotted dashed line. Results correspond to cases A, C, I and Exo at local time 13:00.}
		\label{fig:wt1d} 
\end{center}
\end{figure}

\end{document}